\documentclass[A4, English, nofootinbib,superscriptaddress, manuscript]{revtex4-1}
\usepackage[T1]{fontenc}
\usepackage[latin9]{inputenc}
\usepackage{color}
\usepackage{babel}
\usepackage{array}
\usepackage{multirow}
\usepackage{amsmath}
\usepackage{graphicx}
\usepackage[unicode=true,pdfusetitle,
 bookmarks=true,bookmarksnumbered=false,bookmarksopen=false,
 breaklinks=false,pdfborder={0 0 0},pdfborderstyle={},backref=false,colorlinks=true]
 {hyperref}
\hypersetup{
 linkcolor=blue, citecolor=cyan}

\makeatletter

\providecommand{\tabularnewline}{\\}

\usepackage{amsthm}

\usepackage{xcolor}

\allowdisplaybreaks[2]

\makeatother

\begin{document}
\title{Late-time asymptotic solutions, attractor, and focusing behavior of
spin hydrodynamics}
\author{Dong-Lin Wang}
\email{donglinwang@mail.ustc.edu.cn}

\affiliation{Department of Modern Physics, University of Science and Technology
of China, Anhui 230026, China}
\author{Li Yan }
\email{cliyan@fudan.edu.cn}

\affiliation{Institute of Modern Physics, Fudan University, Shanghai 200433, China}
\author{Shi Pu}
\email{shipu@ustc.edu.cn}

\affiliation{Department of Modern Physics, University of Science and Technology
of China, Anhui 230026, China}
\affiliation{Southern Center for Nuclear-Science Theory (SCNT), Institute of Modern Physics, Chinese Academy of Sciences, Huizhou 516000, Guangdong Province, China}

\begin{abstract}
We have investigated the late-time asymptotic solutions, attractor, and focusing behavior of minimal causal spin hydrodynamics in Bjorken expansion. Using the method of dominant balance, we derive the late-time asymptotic solutions of the evolution equation for spin density and identify the specific conditions necessary for the spin density to exhibit a power-law decay. We then analyze both the late-time and early-time attractors for the decay rate of spin density. Additionally, we report the focusing behavior in spin hydrodynamics, which has not been found in conventional relativistic hydrodynamics in Bjorken expansion. Our findings suggest that spin density can be treated as a conventional hydrodynamic variable at late times under certain conditions.
\end{abstract}
\maketitle

\section{Introduction}

Large orbital angular momentum generated by relativistic heavy ion collisions induces the spin polarization of $\Lambda$ and $\bar{\Lambda}$ hyperons and spin alignment of vector mesons \citep{Liang:2004ph,Liang:2004xn,Gao:2007bc} through the transformation between spin and orbital angular momentum, similar to the Barnett effect \citep{RevModPhys.7.129}. The RHIC-STAR \citep{STAR:2017ckg,STAR:2018gyt,STAR:2019erd,STAR:2020xbm,STAR:2021beb,STAR:2022fan} and LHC-ALICE experiments \citep{ALICE:2019aid,ALICE:2022dyy} have measured the global and local polarization of $\Lambda$ or $\Xi$ hyperons. The global polarization, i.e., the polarization along the direction of initial total angular momentum as a function of collisional energy, can be well understood by various phenomenological models \citep{Becattini:2007nd,Becattini:2007sr,Becattini:2013fla,Fang:2016vpj,Karpenko:2016jyx,Xie:2017upb,Li:2017slc,Sun:2017xhx,Shi:2017wpk,Wei:2018zfb,Xia:2018tes,Vitiuk:2019rfv,Ryu:2021lnx,Lei:2021mvp,Wu:2022mkr,Wu:2023tku}.
However, these models give an opposite sign for the polarization as a function of the azimuthal angle in the transverse plane, i.e., local spin polarization. To understand this discrepancy, shear-induced polarization has been considered \citep{Liu:2021uhn,Fu:2021pok,Becattini:2021suc,Becattini:2021iol,Yi:2021ryh,Yi:2021unq,Florkowski:2021xvy,Alzhrani:2022dpi,Wu:2022mkr,Yi:2023tgg,Wu:2023tku} as well as collisional \cite{Sheng:2021kfc,Wang:2021qnt,Weickgenannt:2021cuo,Fang:2022ttm} and self-energy effects \cite{Fang:2023bbw}. The contribution from electromagnetic fields has also been revisited \cite{Sun:2024isb}. Nevertheless, the problem of local spin polarization has not been fully understood and needs more in-depth studies. For more details and recent progress on spin polarization, we refer to the recent reviews \citep{Gao:2020vbh,Becattini:2022zvf,Hidaka:2022dmn,Becattini:2024uha} and references therein. 


To provide a better understanding of the spin dynamics of quark matter, people have extended the well-established relativistic hydrodynamics to relativistic spin hydrodynamics as a macroscopic effective theory. In addition to the conventional conservation equations for energy-momentum tensor, and currents, spin hydrodynamics provides the evolution equation for spin density, resulting from the conservation of total angular momentum. The formulas for spin hydrodynamics have been constructed by various methods, e.g., kinetic approaches \citep{Bhadury:2020puc,Shi:2020htn,Bhadury:2020cop,Singh:2020rht,Peng:2021ago,Weickgenannt:2022zxs,Weickgenannt:2022jes,Weickgenannt:2022qvh,Wagner:2024fhf}, entropy current analysis \citep{Hattori:2019lfp,Fukushima:2020ucl,Li:2020eon,Gallegos:2021bzp,She:2021lhe,Hongo:2021ona,Cao:2022aku,Hu:2022azy,Biswas:2023qsw}, and other effective theories \citep{Montenegro:2017rbu,Gallegos:2020otk,Garbiso:2020puw}. Moreover, the dependence on pseudo-gauge \citep{Becattini:2018duy,Speranza:2020ilk,Fukushima:2020ucl,Das:2021aar,Buzzegoli:2021wlg,Weickgenannt:2022jes,Daher:2022xon,Singh:2024qvg} and causality-stability conditions \citep{Sarwar:2022yzs,Daher:2022wzf,Xie:2023gbo,Weickgenannt:2023btk,Wang:2023csj,Daher:2024bah,Ren:2024pur} for spin hydrodynamics have also been widely discussed.


Despite much progress, spin hydrodynamics still suffers from several problems. First, it is unclear how large the spin density will be at late times. The linear mode analysis for spin hydrodynamics around a homogeneous equilibrium state (without global rotation) showed that the spin density corresponds to non-hydrodynamic modes \citep{Hattori:2019lfp,Hongo:2021ona,Daher:2022wzf,Xie:2023gbo}, which are damped quickly in the long wavelength limit. Thus, in the framework of spin hydrodynamics proposed in Refs.~\citep{Hattori:2019lfp,Fukushima:2020ucl}, the spin effects may become negligible at late times if the final state is near a homogeneous equilibrium. This naturally calls into question the validity of spin density as a hydrodynamic and thermodynamic variable. Furthermore, if spin hydrodynamics predicts that the spin effects are always negligible at late times, it is doubtful that we should use spin hydrodynamics to describe the quark gluon plasma. However, the linear mode analysis works only near a homogeneous equilibrium state, e.g., see the early pioneer works \citep{Hiscock:1985zz,Hiscock:1987zz} as well as the recent developments in Refs.~\cite{Heller:2022ejw,Gavassino:2023myj,Wang:2023csj,Gavassino:2023mad,Wang:2023csj,Hoult:2023clg}, and the appearance of non-hydrodynamic modes does not mean that the spin density must decay rapidly in other nonequilibrium configurations.
To understand this problem, a straightforward way is to derive the solutions of spin hydrodynamics in such configurations. Although the analytic solutions for spin hydrodynamics in Bjorken \cite{Wang:2021ngp} and Gubser expansion \cite{Wang:2021wqq} have been derived under certain assumptions, systematic studies on it are still lacking.
Second, another more practical problem is that the initial distribution of spin density is unknown when using spin hydrodynamics to study the evolution of spin in relativistic heavy ion collisions. Without suitable initial conditions, the numerical framework for spin hydrodynamics is not well-established.


Fortunately, the recent developments and studies of attractors in conventional relativistic hydrodynamics can help to solve the above problems.
It was found that when the hydrodynamic attractor exists, solutions always converge to the attractor regardless of initial conditions \citep{Heller:2015dha}. Over the past few years, the attractor behavior has been observed in various models \citep{Heller:2016rtz,Romatschke:2017vte,Blaizot:2017ucy,Denicol:2017lxn,Spalinski:2017mel,Strickland:2018ayk,Denicol:2019lio,Blaizot:2019scw,Brewer:2019oha,Almaalol:2020rnu,Blaizot:2020gql,Heller:2020anv,Heller:2020uuy,Noronha:2021syv,Heller:2021oxl,Blaizot:2021cdv,Chen:2022ryi,Brewer:2022ifw,Heller:2023mah,Aniceto:2024pyc}, which potentially unravels the remarkable success of relativistic hydrodynamic simulations in describing the experimental outcomes of heavy-ion collisions \citep{Florkowski:2017olj,Berges:2020fwq,Soloviev:2021lhs,Jankowski:2023fdz}.
A natural question arises: \emph{Does spin hydrodynamics exhibit attractors? If so, what can we learn from their behavior?}


In this work, we will systematically investigate the late-time asymptotic solutions and attractor of the minimal causal spin hydrodynamics in Bjorken expansion. By using the method of dominant balance, we can derive the late-time asymptotic solutions of the evolution equation for spin density. Based on these solutions, we can identify the specific conditions necessary for the spin density to exhibit a power-law decrease over proper time. We will demonstrate that, under certain conditions, there exist both the late-time and early-time attractors for the decay rate of spin density. Additionally, we will report the focusing behavior of spin density, which has never been found in conventional relativistic hydrodynamics in Bjorken expansion.

The paper is organized as follows. In Sec.~\ref{sec:Spin-hydrodynamics},
we briefly review the theoretical framework of the minimal causal spin hydrodynamics. In Sec.~\ref{sec:Spin-hydrodynamics-with},
we derive the equations for energy density and spin density in a Bjorken
flow. In Sec.~\ref{sec:SolveSpinEq}, we study the late-time asymptotic
solutions. In Sec.~\ref{sec:Attractor-behavior},
we discuss the late-time and early-time attractors. In Sec.~\ref{sec:FocusingBehavior},
we report the focusing behavior of spin density. Sec.~\ref{sec:Conclusion-}
concludes this work. 
Throughout this work, we adopt the mostly minus convention for the
Minkowski metric $g_{\mu\nu}=\mathrm{diag}(+,-,-,-)$ and use the projection
tensor $\Delta_{\mu\nu}\equiv g_{\mu\nu}-u_{\mu}u_{\nu}$,
where $u_{\mu}$ denotes fluid velocity. For an arbitrary tensor $A^{\mu\nu}$, we introduce the following notations, 
\begin{eqnarray}
    A^{(\mu\nu)}&\equiv&\frac{1}{2}(A^{\mu\nu}+A^{\nu\mu}), \;\;
    A^{[\mu\nu]}\equiv\frac{1}{2}(A^{\mu\nu}-A^{\nu\mu}),
    \nonumber  \\
A^{<\mu\nu>}&\equiv&\frac{1}{2}(\Delta^{\mu\alpha}\Delta^{\nu\beta}+\Delta^{\nu\alpha}\Delta^{\mu\beta})A_{\alpha\beta}-\frac{1}{3}\Delta^{\mu\nu}(A^{\rho\sigma}\Delta_{\rho\sigma}).
\end{eqnarray}

\section{Brief review on relativistic spin hydrodynamics \label{sec:Spin-hydrodynamics}}

Let us start by a brief review of relativistic spin hydrodynamics. In relativistic spin hydrodynamics, the conservation equations are
\begin{eqnarray}
\nabla_{\mu}\Theta^{\mu\nu} & = & 0,\label{eq:EMConserved}\\
\nabla_{\lambda}J^{\lambda\mu\nu} & = & 0,\label{eq:TAMConserved}
\end{eqnarray}
corresponding to energy momentum conservation and total angular momentum
conservation, respectively. Here, the energy momentum tensor $\Theta^{\mu\nu}$ and the total angular momentum
current tensor $J^{\lambda\mu\nu}$ are expressed in the canonical form \citep{Hattori:2019lfp,Fukushima:2020ucl}, as, 
\begin{eqnarray}
\Theta^{\mu\nu} & = & eu^{\mu}u^{\nu}-(p+\Pi)\Delta^{\mu\nu}+2h^{(\mu}u^{\nu)}+\pi^{\mu\nu}+2q^{[\mu}u^{\nu]}+\phi^{\mu\nu},\label{eq:Thetamunu}\\
J^{\lambda\mu\nu} & = & x^{\mu}\Theta^{\lambda\nu}-x^{\nu}\Theta^{\lambda\mu}+\Sigma^{\lambda\mu\nu},\label{eq:Jlambdamunu}
\end{eqnarray}
where $e$, $p$, $\Pi$, $u^{\mu}$, and $\Sigma^{\lambda\mu\nu}$
represent energy density, thermodynamic pressure, bulk viscous pressure,
fluid velocity, and rank-$3$ spin density current tensor, respectively. The
heat current vector $h^{\mu}$,  shear viscous tensor $\pi^{\mu\nu}$,
and $q^{\mu},\phi^{\mu\nu}$ are defined as 
\begin{eqnarray}
h^{\mu}\equiv u_{\nu}\Theta^{(\mu\nu)}, & \quad  & \pi^{\mu\nu}\equiv\Theta^{<\mu\nu>},\\
q^{\mu}\equiv u_{\nu}\Theta^{[\mu\nu]}, & \quad  & \phi^{\mu\nu}\equiv\Delta^{\mu\alpha}\Delta^{\nu\beta}\Theta_{[\alpha\beta]}.
\end{eqnarray}
Inserting
Eqs.~(\ref{eq:Thetamunu}, \ref{eq:Jlambdamunu})
into Eqs.~(\ref{eq:EMConserved}, \ref{eq:TAMConserved}) yields the evolution equation for the rank-$3$ spin current tensor,
\begin{equation}
\nabla_{\lambda}\Sigma^{\lambda\mu\nu}=-4q^{[\mu}u^{\nu]}-2\phi^{\mu\nu},\label{eq:OriginalSpinEq}
\end{equation}
from which we see that $q^{\mu}$ and $\phi^{\mu\nu}$
serve as the source terms for spin.

The thermodynamic relations for spin hydrodynamics are extended
 \citep{Hattori:2019lfp,Fukushima:2020ucl}
\begin{eqnarray}
e+p & = & sT+\omega_{\mu\nu}S^{\mu\nu},\\
dp & = & sdT+S^{\mu\nu}d\omega_{\mu\nu},
\end{eqnarray}
where we have neglected the term for particle number density. 
The local spin density tensor $S^{\mu\nu}\sim\mathcal{O}(1)$
is defined as \citep{Fukushima:2020ucl} 
\begin{eqnarray}
\Sigma^{\lambda\mu\nu} & = & u^{\lambda}S^{\mu\nu}+\mathcal{O}(\partial).
\end{eqnarray}
The quantity $\omega_{\mu\nu}\sim\mathcal{O}(\partial)$, conjugate
to $S^{\mu\nu}$, is called spin chemical potential.

For simplicity, in this work we concentrate on the minimal causal spin hydrodynamics \citep{Xie:2023gbo} with $h^\mu=0$, and neglect other second-order
terms. For a complete second-order
theory of spin hydrodynamics, one can refer to Refs.~\citep{Weickgenannt:2022zxs,Biswas:2023qsw}.
The constitutive relations in the minimal causal spin hydrodynamics 
are given by \citep{Liu:2020ymh,Xie:2023gbo} 
\begin{eqnarray}
\tau_{q}\Delta^{\mu\nu}u^{\alpha}\nabla_{\alpha}q_{\nu}+q^{\mu} & = & \lambda(T^{-1}\Delta^{\mu\alpha}\nabla_{\alpha}T+u^{\alpha}\nabla_{\alpha}u^{\mu}-4\omega^{\mu\nu}u_{\nu}),\label{eq:Const1}\\
\tau_{\phi}\Delta^{\mu\alpha}\Delta^{\nu\beta}u^{\rho}\nabla_{\rho}\phi_{\alpha\beta}+\phi^{\mu\nu} & = & 2\gamma\Delta^{\mu\alpha}\Delta^{\nu\beta}(\nabla_{[\alpha}u_{\beta]}+2\omega_{\alpha\beta}),\label{eq:Const2}\\
\tau_{\pi}\Delta^{\alpha<\mu}\Delta^{\nu>\beta}u^{\rho}\nabla_{\rho}\pi_{\alpha\beta}+\pi^{\mu\nu} & = & 2\eta\nabla^{<\mu}u^{\nu>},\label{eq:Const3}\\
\tau_{\Pi}u^{\rho}\nabla_{\rho}\Pi+\Pi & = & -\zeta\nabla_{\mu}u^{\mu},\label{eq:Const4}
\end{eqnarray}
with the relaxation times $\tau_{q},\tau_{\phi},\tau_{\pi},\tau_{\Pi}>0$
and transport coefficients $\lambda,\gamma,\eta,\zeta>0$. 
As shown in Ref.~\citep{Xie:2023gbo},
the constitutive relations (\ref{eq:Const1}-\ref{eq:Const4}) correspond
to causal spin hydrodynamic equations. 

\section{Relativistic spin hydrodynamics with Bjorken expansion \label{sec:Spin-hydrodynamics-with}}

In the early stages of relativistic heavy ion collisions, the quark-gluon plasma
 expands rapidly along the longitudinal direction ($z$-axis in the laboratory frame), which is roughly boost invariant, while the transverse expansion ($xy$-plane) is almost negligible. A well-known description for this stage is the Bjorken
expansion \citep{Bjorken:1982qr}, which according to the Bjorken symmetry gives rise to the velocity profile, 
\begin{equation}
u^{\mu}=(u^{\tau},u^{x},u^{y},u^{\xi})=(1,0,0,0),\label{eq:BjorkenVelo}
\end{equation}
in the Milne coordinates $(\tau,x,y,\xi)$. Here $\tau$ denotes
the longitudinal proper time, and $\xi$ represents the space-time rapidity,
\begin{equation}
\tau\equiv\sqrt{t^{2}-z^{2}},\quad\xi\equiv\frac{1}{2}\ln\frac{t+z}{t-z}.
\end{equation}

With respect to the Bjorken expansion, the system is highly symmetric that
conformal fluid dynamics can be analytically solved. In the case of the
relativistic spin hydrodynamics, following the similar strategy, we first 
assume 
that $\Pi=0$ and $\pi^{\mu\nu}=0$, while $q^{\mu}$ and $\phi^{\mu\nu}$ can be nonvanishing. 
It is equivalent to the hierarchy of the relaxation times: $\tau_\phi\sim \tau_q \gg \tau_\pi \sim \tau_\Pi$, following the fact that exchange of energy and momentum is overwhelming strong than that of spin via interactions, e.g. see Ref.~\cite{Fang:2022ttm} the calculations from the quantum kinetic theory in hard thermal loop approximation.
To close the equations of motion, we consider the linearized relations for the 
pressure and the spin chemical potential, as a generalization of the conformal equation of state,
\begin{equation}
p=c_{s}^{2}e,\quad 
S^{\mu\nu}=\chi\omega^{\mu\nu}+2\chi^{\prime}u_{\lambda}\omega^{\lambda[\mu}u^{\nu]},
\label{eq:Eos2}
\end{equation}
where the speed of sound $c_{s}=1/\sqrt{3}$ is treated as a constant, and the $\chi, \chi^\prime$ are two spin susceptibilities satisfying 
$\chi^{\prime}>\chi>0$ 
\citep{Xie:2023gbo,Daher:2024ixz}.  

Third, we assume that, in the Milne coordinates
$(\tau,x,y,\xi)$, the quantities $e$, $p$, $q^{\mu}$, and $S^{\mu\nu}$
depend only on $\tau$ and are independent of $x,y,\xi$. 
We then insert the velocity profile in Eq.~(\ref{eq:BjorkenVelo}) into the
conservation Eqs.~(\ref{eq:EMConserved}, 
\ref{eq:TAMConserved}) and the constitutive relations in Eqs.~(\ref{eq:Const1}-\ref{eq:Const4}) as in our previous work \citep{Wang:2021ngp}. 
More details can be found in  Appendix. \ref{sec:Equations-in-Bjorken}.

The energy conservation equation becomes 
\begin{equation}
\frac{de}{d\tau}+(1+c_{s}^{2})\frac{e}{\tau}=0,\label{eq:EnergyDensityEq}
\end{equation}
whose solution is
\begin{equation}
e=e_{0}\left(\frac{\tau_{0}}{\tau}\right)^{1+c_{s}^{2}}=e_{0}\left(\frac{\tau_{0}}{\tau}\right)^{4/3},\label{eq:Sol_e}
\end{equation}
with $\tau_{0}$ being the initial time and $e_{0}$ being the initial
energy density. Note that under our assumption, the spin density tensor or spin chemical potential do not contribute to the solution (\ref{eq:Sol_e}).

Now, let us focus on the evolution equation for spin density,
\begin{equation}
\tau_{\phi}^{2}\frac{d^{2}S}{d\tau^{2}}+\left(\frac{\tau_{\phi}}{\tau}+1\right)\tau_{\phi}\frac{dS}{d\tau}+\left[\frac{\tau_{\phi}}{\tau}-\left(\frac{\tau_{\phi}}{\tau}\right)^{2}+8\frac{\tau_{\phi}\gamma}{\chi}\right]S=0,\label{eq:SpinDensityEq}
\end{equation}
where 
\begin{equation}
S\in\{S^{xy},\tau S^{x\xi},\tau S^{y\xi}\}.\label{eq:NonzeroS}
\end{equation}
The spin components $S^{\tau x}$, $S^{\tau y}$, and $S^{\tau\xi}$,
need to be zero under the aforementioned assumptions. Otherwise,
the self-consistent solutions do not exist, also see the discussion in Appendix. \ref{sec:Equations-in-Bjorken}.
Returning to the conventional $(t,x,y,z)$ coordinates, we obtain
\begin{eqnarray}
S^{tx}=-(\tau S^{x\xi})\sinh\xi, & \  & S^{xz}=(\tau S^{x\xi})\cosh\xi,\label{eq:Spin_txyz_1}\\
S^{ty}=-(\tau S^{y\xi})\sinh\xi, & \  & S^{yz}=(\tau S^{y\xi})\cosh\xi,\label{eq:Spin_txyz_2}\\
S^{tz}=0, & \  & S^{xy}=S^{xy}.\label{eq:Spin_txyz_3}
\end{eqnarray}

We obtain four additional non-vanishing components of the spin density tensor, 
besides $S^{xy}$ \citep{Wang:2021ngp}.
However, we find that the other four components $S^{ty},S^{tx},S^{yz}$,
and $S^{xz}$ (or equivalently $S^{x\xi}$ and $S^{y\xi}$) should
vanish to satisfy the symmetry constraints in Bjorken expansion. 
The symmetry transformations of Bjorken expansion are : 
(1) translation in the $xy$-plane, (2) rotation along the $z$-direction, (3)  boost along the $z$-direction \citep{Bjorken:1982qr,Gubser:2010ze}.
Imposing these symmetries on the spin density $S^{\mu\nu}$, we obtain
\citep{Gubser:2010ze}
\begin{equation}
0=\mathcal{L}_{\kappa}S^{\mu\nu}=\kappa^{\lambda}\partial_{\lambda}S^{\mu\nu}-\partial_{\lambda}\kappa^{\mu}S^{\lambda\nu}-\partial_{\lambda}\kappa^{\nu}S^{\mu\lambda},\label{eq:LkSmunu}
\end{equation}
where $\mathcal{L}_{\kappa}$ is the Lie derivative along $\kappa$,
and $\kappa$ represents the generators for the aforementioned symmetry
transformations, i.e., 
\begin{equation}
\kappa\in\left\{ \partial_{x},\ \partial_{y},\ x\partial_{y}-y\partial_{x},\ z\partial_{t}+t\partial_{z}\right\} .\label{eq:generators}
\end{equation}
Eq.~(\ref{eq:LkSmunu}) leads to  
\begin{eqnarray}
\partial_{x}S^{\mu\nu}=\partial_{y}S^{\mu\nu} & = & 0,\label{eq:Stranslation}\\
S^{ty}=S^{tx}=S^{yz}=S^{xz} & = & 0,\label{eq:Srotation}\\
\partial_{\xi}S^{xy}=\partial_{\xi}S^{tz} & = & 0,\label{eq:Sboost}
\end{eqnarray}
from which we find that the only nonzero spin components can be $S^{xy}$
and $S^{tz}$. But the self-consistency of hydrodynamic equations
demands $S^{tz}=0$. Therefore, we have only one nonzero component
$S^{xy}$.

\section{Late-time asymptotic solutions \label{sec:SolveSpinEq}}

Thermalization of the spin degrees of freedom in the system is characterized by
Eq.~(\ref{eq:SpinDensityEq}). 
In general, the relaxation time $\tau_\phi$ can be a function of proper time $\tau$. We then introduce the inverse Knudsen number $\mathrm{Kn}^{-1}$ as, 
\begin{equation}
\mathrm{Kn}^{-1}\approx w\equiv\frac{\tau}{\tau_{\phi}}.\label{eq:Knudsen_number}
\end{equation}
We follow Ref.~\citep{Blaizot:2021cdv} to parameterize it as, 
\begin{equation}
w\equiv\frac{\tau}{\tau_{\phi}}=\left(\frac{\tau}{\tau_{1}}\right)^{\Delta_{1}},\ \frac{\tau_{\phi}\gamma}{\chi}=\alpha w^{\Delta_{2}},\label{eq:Kn-t}
\end{equation}
where $\alpha$ and $\Delta_{1,2}$ are dimensionless constants, with
$\alpha>0$ and $\Delta_{1}\neq0$.
The constant $\tau_{1}$ serves as a measure of time at
which the collision rate equals the expansion rate, since $1/\tau_{\phi}$
and $1/\tau=\nabla_{\mu}u^{\mu}$ represent the collision rate and
volumetric expansion rate, respectively. 
When $w$ or the Knudsen number is a constant, the parameterization in Eq.~(\ref{eq:Kn-t}) is inappropriate, we also study this special case and present the discussion and results in Appendix \ref{sec:AsymptoticS_C}.
Let us take a close look at the parameter $\Delta_2$. For a constant spin susceptibility $\chi$, according to Eq.~(\ref{eq:Const2}), $\Delta_2$ describes the enhancement of the response of the source term $\phi^{\mu\nu}$ to $\Delta^{\mu\alpha}\Delta^{\nu\beta}(\nabla_{[\alpha}u_{\beta]} + 2\omega_{\alpha\beta})$. When $w \rightarrow \infty$, a larger $\Delta_2$ indicates a greater response of the source term $\phi^{\mu\nu}$, i.e., a greater variation of spin density induced accordingly.

In this work, we focus on the cases where $\Delta_{1}>0$, i.e. we discuss the small
$\mathrm{Kn}$ case when $\tau\rightarrow+\infty$. It indicates that the
collision rate overcomes volumetric expansion rate, leading the system
into a hydrodynamic regime at late times. It also ensures
that the long-time evolution of Eq.~(\ref{eq:S_w}) is well-defined. Conversely, for $\Delta_{1}<0$,
we have large $\mathrm{Kn}$ as $\tau\rightarrow+\infty$,
corresponding to a free streaming regime where the volumetric expansion
dominates. In this limit, the gradient expansion scheme
fails and the hydrodynamic equations cannot describe the system well. 



Using the parameterization in Eq.~(\ref{eq:Kn-t}), we rewrite Eq.~(\ref{eq:SpinDensityEq}) as 
\begin{equation}
\frac{d^{2}S^{xy}}{dw^{2}}+(\Delta_{1}^{-1}+w^{-1})\frac{dS^{xy}}{dw}+\Delta_{1}^{-2}(w^{-1}-w^{-2}+8\alpha w^{\Delta_{2}})S^{xy}=0.\label{eq:S_w}
\end{equation}
In analogy to the relative decay rate of energy density in the description of pre-equilibrium evolution of ordinary fluid, 
we introduce the relative decay rate of the spin density,
\begin{equation}
f(w)\equiv\Delta_{1}\frac{w}{S^{xy}}\frac{dS^{xy}}{dw}=\frac{\tau}{S^{xy}}\frac{dS^{xy}}{d\tau},\label{eq:fS}
\end{equation}
and Eq.~(\ref{eq:S_w}) becomes a first order nonlinear ordinary differential equation, 
\begin{equation}
\Delta_{1}wf^{\prime}+f^{2}+wf+w-1+8\alpha w^{2+\Delta_{2}}=0,\label{eq:RewriteSpinEvo1}
\end{equation}
where $f^{\prime}=df/dw$. 
It should be aware that, unlike the energy density in the ordinary fluid which is positive by definition, the spin density in the medium may vanish with respect to the conservation of total angular momentum.

The general solution of Eq.~(\ref{eq:S_w}) is given by 
a linear combination of two independent solutions $S_{(1)}$ and $S_{(2)}$,
i.e., 
\begin{equation}
S^{xy}(w;c_{(1)},c_{(2)})\equiv c_{(1)}S_{(1)}(w)+c_{(2)}S_{(2)}(w),\label{eq:LineaeComb}
\end{equation}
where $c_{(1)}$ and $c_{(2)}$ are integration
constants determined by initial condition. 
Then general solutions of Eq.~(\ref{eq:RewriteSpinEvo1}), denoted by $f(w;C)$ with an integration constant $C$, 
accordingly have the form\footnote{If $c_{(2)}=0$, Eq.~(\ref{eq:fwC}) should be understood as $\lim_{C\rightarrow\infty}f(w;C)=\Delta_{1}wS_{(1)}^{\prime}(w)/S_{(1)}(w)$.} 
\begin{equation}
f(w;C)=\Delta_{1}w\frac{CS_{(1)}^{\prime}(w)+S_{(2)}^{\prime}(w)}{CS_{(1)}(w)+S_{(2)}(w)},\label{eq:fwC}
\end{equation}
where $C=c_{(1)}/c_{(2)}$. 
If we have already known two special solutions $f(w;C_{1})$ and $f(w;C_{2})$, then we can integrate Eq.~(\ref{eq:fS}) and obtain $S^{xy}(w;C_{1},1)$
and $S^{xy}(w;C_{2},1)$, which are linearly independent solutions
of Eq.~(\ref{eq:S_w}) provided $C_{1}\neq C_{2}$. 

We compute the asymptotic solutions of Eq.~(\ref{eq:RewriteSpinEvo1})
by using the method of dominant balance \citep{bender1999advanced,paulsen2013asymptotic}.
In asymptotic analysis, dominant balance means selecting the terms that are asymptotically larger than others and, then 
solving the equations given by the selected terms only.
Here, we derive the leading order
solutions for the special case $\Delta_{2}=0$ as an example. Other asymptotic solutions can be obtained following the same strategy.  
If $|f|\ll w$ as
$w\rightarrow+\infty$, the term $8\alpha w^{2}$ dominates 
so that Eq.~(\ref{eq:RewriteSpinEvo1}) cannot be balanced.
Similarly, if $|f|\gg w$ as $w\rightarrow+\infty$, the term $f^{2}$
is much larger than other terms and Eq.~(\ref{eq:RewriteSpinEvo1})
cannot be balanced. Hence, the leading order of $f$, denoted as $f_{0}$,
should be proportional to $w$. Then Eq.~(\ref{eq:RewriteSpinEvo1}) reduces to, 
\begin{equation}
f_{0}^{2}+wf_{0}+8\alpha w^{2}=0,
\end{equation}
which gives $f_{0}=w(-1\pm\sqrt{1-32\alpha})/2$. For the next-leading
order, we can substitute $f=f_{0}+f_{1}$ into Eq.~(\ref{eq:RewriteSpinEvo1})
and derive the leading order of $f_{1}$ with the same strategy. 

We summarize the asymptotic solutions in
Table~\ref{tab:AsympSolLarge}. The symbols $\propto$ and $\sim$
used in Table \ref{tab:AsympSolLarge} are defined as follows, 
\begin{eqnarray}
A(w)\sim e^{B(w)}\textrm{ as }w\rightarrow+\infty & \Leftrightarrow & \lim_{w\rightarrow+\infty}A(w)/e^{B(w)}=1,\label{eq:AAsympto}\\
A(w)\propto e^{B(w)}\textrm{ as }w\rightarrow+\infty & \Leftrightarrow & \lim_{w\rightarrow+\infty}\ln A(w)/B(w)=1,\label{eq:APropto}
\end{eqnarray}
where $A(w)$ and $B(w)$ stand for two functions of $w$. The definition in Eq.~(\ref{eq:APropto}) implies that if $A(w)\propto e^{-aw^{b}}$ with
constants $a,b>0$, then $A(w)/e^{-(a+\epsilon)w^{b}}\rightarrow+\infty$
as $w\rightarrow+\infty$ for any positive constant $\epsilon>0$.
Generally, $A(w)\sim e^{B(w)}$ implies $A(w)\propto e^{B(w)}$ but
the reverse is not true. For example, for $A(w)=e^{-w^{2}+w}$ and
$B(w)=-w^{2}$, we can write $A(w)\propto e^{B(w)}$ but $A(w)\sim e^{B(w)}$
is false. 

\begin{table}
\caption{Asymptotic solutions of Eq.~(\ref{eq:S_w}) as $w\rightarrow+\infty$.
Here $S_{(1)}$ and $S_{(2)}$ represent two linearly independent
solutions of Eq.~(\ref{eq:S_w}).  The symbols $\propto$ and $\sim$ used 
here are defined in Eqs.~(\ref{eq:AAsympto}, \ref{eq:APropto}).
 \label{tab:AsympSolLarge}}

\noindent \centering{}%
\begin{tabular}{c|ccc}
\cline{1-3} \cline{2-3} \cline{3-3} 
\noalign{\vskip0.2cm}
\multicolumn{3}{c}{$w\rightarrow+\infty$} & \tabularnewline[0.2cm]
\cline{1-3} \cline{2-3} \cline{3-3} 
\multirow{2}{*}{$\Delta_{2}>0$} &  & \multirow{2}{*}{$S_{(1),(2)}\propto e^{-w/(2\Delta_{1})}$} & \multirow{2}{*}{}\tabularnewline
 &  &  & \tabularnewline
\cline{1-3} \cline{2-3} \cline{3-3} 
\multirow{2}{*}{$\Delta_{2}=0$} &  & \multirow{2}{*}{$S_{(1),(2)}\propto e^{-w(1\pm\sqrt{1-32\alpha})/(2\Delta_{1})}$} & \tabularnewline
 &  &  & \tabularnewline
\cline{1-3} \cline{2-3} \cline{3-3} 
\multirow{2}{*}{$-1<\Delta_{2}<0$} &  & \multirow{2}{*}{$S_{(1)}\propto e^{-w/\Delta_{1}}$, $S_{(2)}\propto\exp\left[-\frac{8\alpha w^{1+\Delta_{2}}}{\Delta_{1}(1+\Delta_{2})}\right]$} & \tabularnewline
 &  &  & \tabularnewline
\cline{1-3} \cline{2-3} \cline{3-3} 
\multirow{2}{*}{$\Delta_{2}=-1$} &  & \multirow{2}{*}{$S_{(1)}\propto e^{-w/\Delta_{1}}$, $S_{(2)}\sim w^{-(1+8\alpha)/\Delta_{1}}$} & \tabularnewline
 &  &  & \tabularnewline
\cline{1-3} \cline{2-3} \cline{3-3} 
\multirow{2}{*}{$\Delta_{2}<-1$} &  & \multirow{2}{*}{$S_{(1)}\propto e^{-w/\Delta_{1}}$, $S_{(2)}\sim w^{-1/\Delta_{1}}$} & \tabularnewline
 &  &  & \tabularnewline
\cline{1-3} \cline{2-3} \cline{3-3} 
\end{tabular}
\end{table}

If $\Delta_{2}>-1$, the linearly independent solutions $S_{(1)}$
and $S_{(2)}$ are both controlled by exponentially decaying factors,
as shown in the first three rows of Table. \ref{tab:AsympSolLarge}.
In this case, all the solutions for spin density must be exponentially
decaying, different from the energy density $e$ as shown in Eq.~(\ref{eq:Sol_e}). 
If $\Delta_{1}$ is not very large, suggesting that the spin density decays rapidly, then the spin effects can be negligible at late times.

If $\Delta_{2}\leq-1$, we observe that the spin density can be power-law decaying.
In the last two rows of Table.~\ref{tab:AsympSolLarge}, we find that
one solution, denoted by $S_{(1)}$, exponentially decays, while the
other, $S_{(2)}$, exhibits power-law decay. Recalling that the general
solution is a linear combination of $S_{(1)}$ and $S_{(2)}$, as
shown in Eq.~(\ref{eq:LineaeComb}), the contribution of $S_{(1)}$ becomes negligible compared to $S_{(2)}$ as $w\rightarrow+\infty$.
Consequently, the spin density $S^{xy}$ undergoes
power-law decay at late times, provided $c_{(2)}\neq0$. 
Later, our results will show that numerous initial conditions satisfy the inequality $c_{(2)}\neq0$.

\begin{figure}
\noindent \begin{centering}
\includegraphics[scale=0.6]{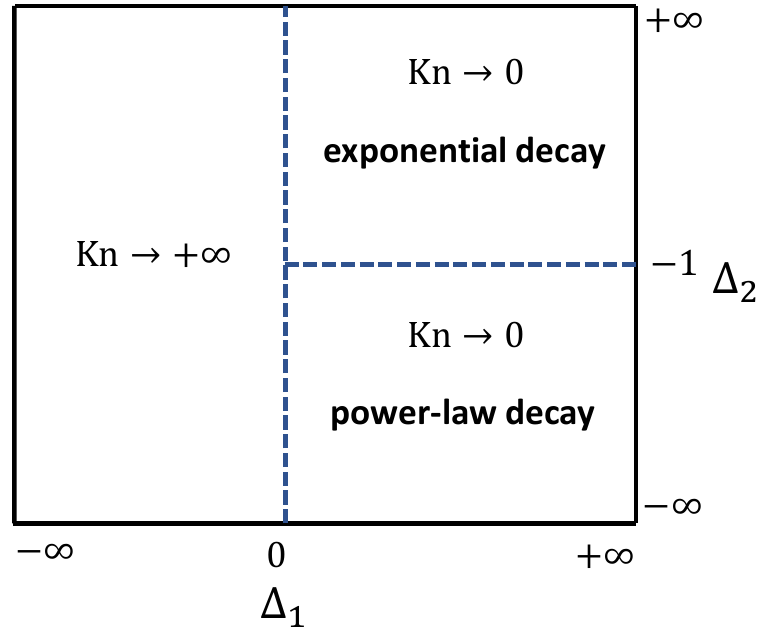}
\par\end{centering}
\noindent \centering{}\caption{The Knudsen number and decaying behavior of spin
density at late time depend on the values of $\Delta_{1,2}$. \label{fig:S0Sp}}
\end{figure}

We summarize the Knudsen number and decaying behavior of spin density at late times for different values $\Delta_{1,2}$ in Fig.~\ref{fig:S0Sp}.
We notice that the hydrodynamic variables in Bjorken flow, such as energy density $e$ 
in Eq.~(\ref{eq:Sol_e}), undergo power-law decay instead of exponential decay. Therefore, we next discuss the cases where 
spin density, as a hydrodynamic variable, is considered when $\Delta_{1,2}$
lie within the domain 
\begin{equation}
\Delta_{1}>0,\quad\Delta_{2}\leq-1.\label{eq:DomainSpinHydro}
\end{equation}

Let us now discuss the constraints imposed by the inequalities (\ref{eq:DomainSpinHydro}). Generally, the parameters $\tau_{\phi}$,
$\gamma$, and $\chi$ depend on the mass $m$ of single particle
and the temperature $T$ \footnote{Since $S^{\mu\nu}\omega_{\mu\nu}\sim\mathcal{O}(\partial)$, 
the spin term does not contribute to the following leading-order analysis.
We also assume that the charge and axial
chemical potentials can be neglected.}. The dimensional analysis leads to 
\begin{eqnarray}
\tau_{\phi} & \approx & \mathrm{const}\cdot m^{\delta_{1}}T^{-1-\delta_{1}}\approx\mathrm{const}\cdot\tau^{(1+\delta_{1})/3},\label{eq:DimensionAna01}\\
\frac{\tau_{\phi}\gamma}{\chi} & \approx & \mathrm{const}\cdot m^{\delta_{2}}T^{-\delta_{2}}\approx\mathrm{const}\cdot\tau^{\delta_{2}/3},\label{eq:DimensionAna02}
\end{eqnarray}
where $\delta_{1,2}$ are dimensionless constants. In the second step
of Eqs.~(\ref{eq:DimensionAna01}, \ref{eq:DimensionAna02}), we have
used $T\approx\mathrm{const}\cdot e^{1/4}$ and Eq.~(\ref{eq:Sol_e}).
By comparing Eqs.~(\ref{eq:DimensionAna01}, \ref{eq:DimensionAna02})
with Eq.~(\ref{eq:Kn-t}), we obtain
\begin{equation}
\Delta_{1}=\frac{1}{3}(2-\delta_{1}),\quad\Delta_{2}=\frac{\delta_{2}}{2-\delta_{1}}.\label{eq:Delta12Expr}
\end{equation}
The inequality $\Delta_{1}>0$ necessitates
\begin{equation}
\delta_{1}<2,
\end{equation}
while $\Delta_{2}\leq-1$ yields 
\begin{equation}
\delta_{2}\leq\delta_{1}-2<0.\label{eq:Constranit2}
\end{equation}
It is worth noting that we must set $\delta_{1,2}=0$ in Eq.~(\ref{eq:DimensionAna01})
if $m=0$, contradicting Eq.~(\ref{eq:Constranit2}). 
In the massless case,  the dimensional analysis in Eqs.~(\ref{eq:DimensionAna01},\ref{eq:DimensionAna02}) needs to be extended to the cases
involving the charge and axial chemical potentials in general.

\section{Attractor \label{sec:Attractor-behavior}}

\subsection{Late-time attractor \label{subsec:LateTime}}

As discussed in the previous section, when $\Delta_{2}\leq-1$,
all the solutions for spin density $S^{xy}$ can be expressed as a
linear combination of an exponential decay solution $S_{(1)}$ and
a power-law decay solution $S_{(2)}$. 
We introduce auxiliary functions $f_{(1),(2)}$ corresponding
to $S_{(1),(2)}$, respectively. From Eq.~(\ref{eq:fS}), we find that, 
when $w \rightarrow \infty$
\begin{eqnarray}
    f_{(1)} &\rightarrow& - w, \\
    f_{(2)} &\rightarrow& \begin{cases}
        -1-8\alpha, & \Delta_{2}=-1,\\
        -1, & \Delta_{2}<-1,
        \end{cases} 
\end{eqnarray}
This implies that there exists a late-time attractor for $f$ when
$\Delta_{2}\leq-1$. In Appendix. \ref{sec:TransseriesSol}, we also
confirm this observation by studying the transseries solutions of
 Eq.~(\ref{eq:RewriteSpinEvo1}). 

\begin{figure}
\noindent \begin{centering}
\includegraphics[scale=0.5]{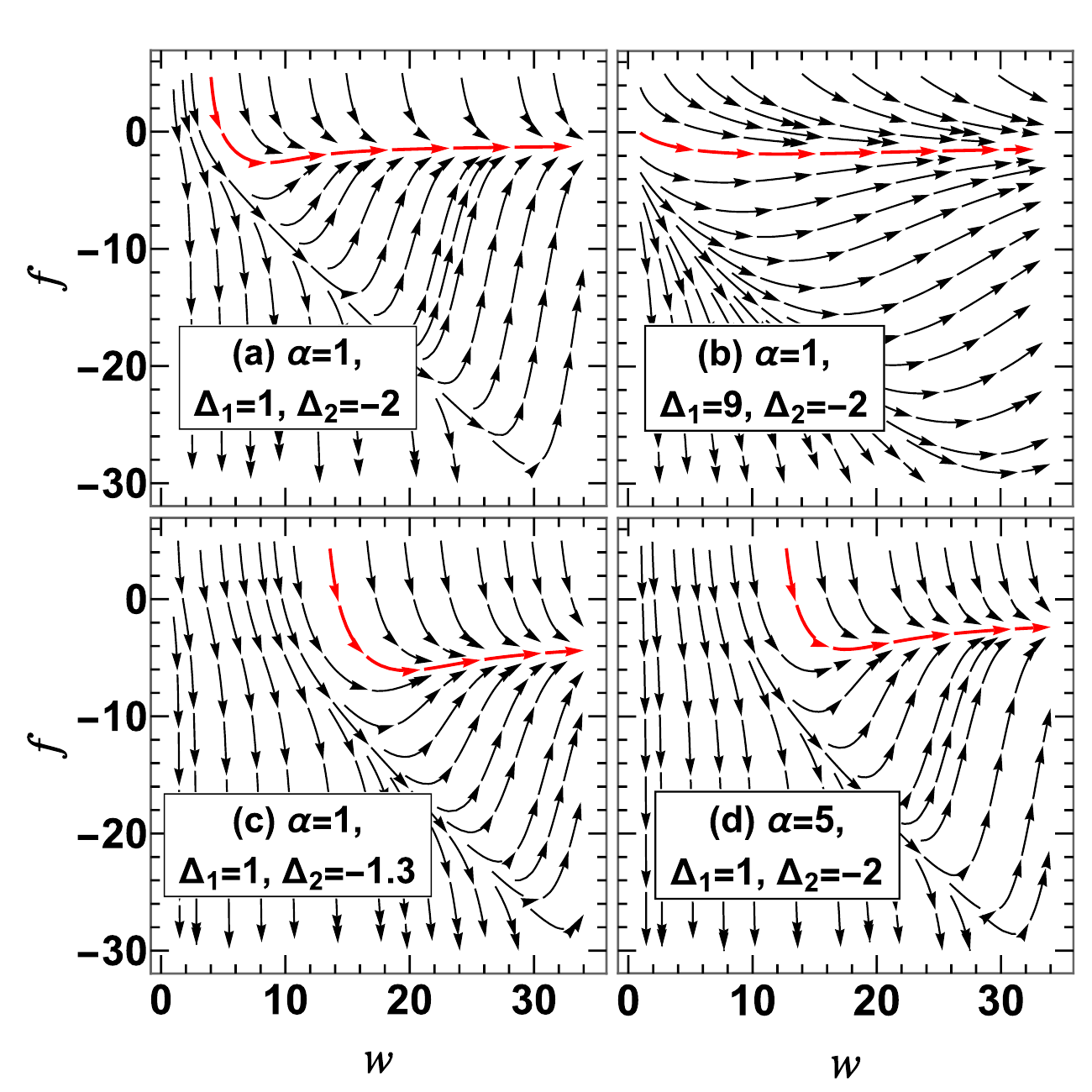}
\par\end{centering}
\noindent \centering{}\caption{Streamlines for the Eq.~(\ref{eq:RewriteSpinEvo1})
with the following parameter sets: (a) $\alpha=1$, $\Delta_{1}=1$,
$\Delta_{2}=-2$; (b) $\alpha=1$, $\Delta_{1}=9$, $\Delta_{2}=-2$;
(c) $\alpha=1$, $\Delta_{1}=1$, $\Delta_{2}=-1.3$; (d) $\alpha=5$,
$\Delta_{1}=1$, $\Delta_{2}=-2$. The arrows of the streamlines indicate
the direction where $w$ increases. The red lines represent the late-time
attractor determined numerically. \label{fig:StreamlinesLate}}
\end{figure}

To provide a clearer insight, we depict the streamlines for 
 Eq.~(\ref{eq:RewriteSpinEvo1})
with $\Delta_{2}<-1$ in Fig.~\ref{fig:StreamlinesLate}. We observe
that solutions around $f_{(2)}$ will
eventually tend to $f_{(2)}$, indicating an attractor behavior
for $f$. On the contrary, solutions near $f_{(1)}$ end up
moving away from $f_{(1)}$, demonstrating a repeller behavior
for $f$. This phenomenon can be understood by analyzing the stability of the
solutions $f_{(1),(2)}$ . We substitute $f=f_{(i)}+\delta f_{(i)}$,
$i=1,2$, into Eq.~(\ref{eq:RewriteSpinEvo1}), and find that 
\begin{eqnarray}
\delta f_{(1)} & \propto & e^{w/\Delta_{1}}\gg1, \nonumber \\
\delta f_{(2)} & \propto & e^{-w/\Delta_{1}}\ll1,
\end{eqnarray}
as $w\rightarrow+\infty$. It means that $f_{(2)}$ is stable while $f_{(1)}$
is unstable, consistent with the streamlines in Fig.~\ref{fig:StreamlinesLate}. 

Let us now discuss the influence of parameters $\alpha$ and $\Delta_{1,2}$
on the formation of the late-time attractor. We rewrite Eq.~(\ref{eq:RewriteSpinEvo1}) as
\begin{equation}
f^{\prime}=-\frac{1}{\Delta_{1}w}\left(f^{2}+wf+w-1+8\alpha w^{2+\Delta_{2}}\right).\label{eq:fprime_fw}
\end{equation}
A key observation is that a smaller $\Delta_{1}$ results in a larger
absolute slope $|f^{\prime}|$. Hence, streamlines around the attractor
converge more rapidly as $\Delta_{1}$ decreases, illustrated in
Fig.~\ref{fig:StreamlinesLate}(a) and (b). Eq.~(\ref{eq:fprime_fw})
also implies that, with sufficiently large $\alpha$ and $\Delta_{2}$,
the slope $f^{\prime}$ becomes highly negative for small $w$, preventing
convergence to the attractor in the small $w$ regime. This phenomenon
can be found in Fig.~\ref{fig:StreamlinesLate}(c) and (d),
where the attractor behavior manifests only for $w>16$. Additionally,
we notice that the convergence of the
late-time attractor
toward $f_{(2)} \rightarrow -1$ becomes slower as $\alpha$ and $\Delta_{2}$ increase, e.g. see red lines in Fig.~\ref{fig:StreamlinesLate}(c) and (d).
In summary, the late-time attractor behavior becomes more evident
for smaller values of $\alpha$ and $\Delta_{1,2}$,
where $\alpha,\Delta_{1}>0$, and $\Delta_{2}\leq-1$. Here,  small $\alpha$ and $\Delta_{2}$ means that the dimensionless quantity $\tau_{\phi}\gamma/\chi$ is small for $w>1$.

\subsection{Early-time attractor \label{subsec:Early01}}

As discussed in Sec. \ref{sec:SolveSpinEq}, the hydrodynamic
equations are valid for small Knudsen number, corresponding to large
$w$ for $\Delta_{1}>0$. During the early-time evolution, characterized
by large Knudsen numbers, the system is far from equilibrium, necessitating
the use of other effective theories, e.g. kinetic theory, for an accurate description. Studies on
kinetic theory without spin effects at early times have revealed the existence
of early-time attractor behavior, and the characteristics of this
early-time attractor can be qualitatively described by second-order
hydrodynamics \citep{Heller:2016rtz,Romatschke:2017vte,Behtash:2017wqg,Blaizot:2017ucy,Strickland:2018ayk,Kurkela:2019set,Almaalol:2020rnu,McNelis:2020jrn,Kamata:2020mka,Blaizot:2020gql,Jaiswal:2022udf}.
As a systematic study, here we also study the early-time attractor of the
minimal causal second order spin hydrodynamics. 

\begin{figure}
\noindent \begin{centering}
\includegraphics[scale=0.5]{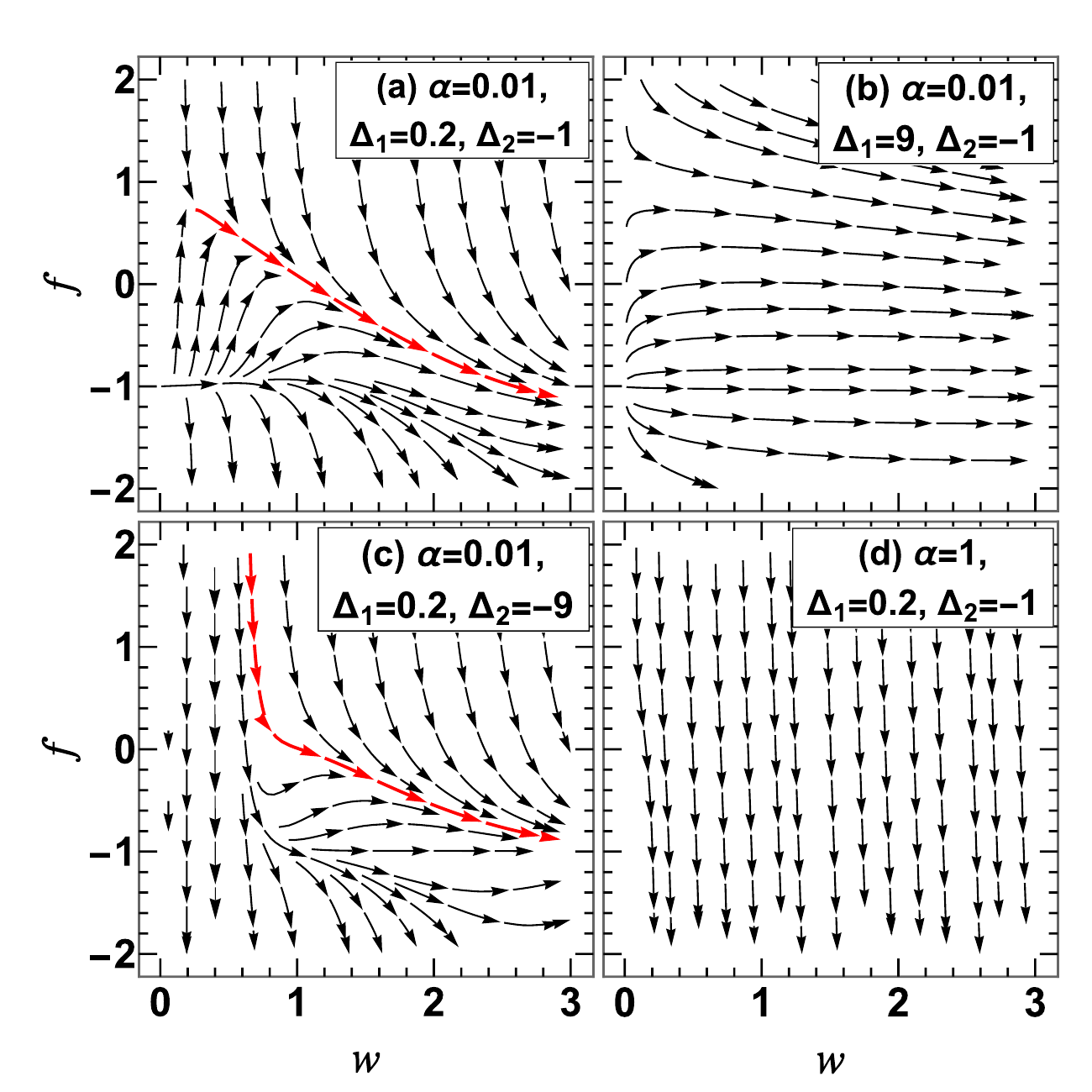}
\par\end{centering}
\caption{Streamlines for the Eq.~(\ref{eq:RewriteSpinEvo1}) 
with the following parameter sets: (a) $\alpha=0.01$, $\Delta_{1}=0.2$,
$\Delta_{2}=-1$; (b) $\alpha=0.01$, $\Delta_{1}=9$, $\Delta_{2}=-1$;
(c) $\alpha=0.01$, $\Delta_{1}=0.2$, $\Delta_{2}=-9$; (d) $\alpha=1$,
$\Delta_{1}=0.2$, $\Delta_{2}=-1$. The arrows of the streamlines
indicate the direction where $w$ increases. The red lines represent
the early-time attractor determined numerically. We do not find an
apparent attractor in the panels (c) and (d). \label{fig:StreamlinesEarly}}
\end{figure}

Due to the complexity of the Eq.~(\ref{eq:S_w}) when $w \rightarrow 0$, we only present the numerical results for the early time attractors here.
We plot the streamlines for Eq.~(\ref{eq:RewriteSpinEvo1}) with different
parameter sets in Fig.~\ref{fig:StreamlinesEarly}. Comparing Fig.~\ref{fig:StreamlinesEarly}(a) and (c), it is apparent
that a smaller $\Delta_{2}$ corresponds to a more rapid decline at
$w<1$, resulting in the absence of attractor behavior. This contrasts
with the cases of a late-time attractor, where a smaller $\Delta_{2}$
enhances the visibility of the attractor.  Our explanation is as follows.
Assuming $f$, $\alpha$ and $w$ in
Eq.~(\ref{eq:fprime_fw}) are fixed, as $\Delta_{2}$ decreases, the
last term $8\alpha w^{2+\Delta_{2}}$ becomes larger if $w<1$, while
$8\alpha w^{2+\Delta_{2}}$ becomes smaller if $w>1$. Since a significantly
large $8\alpha w^{2+\Delta_{2}}$ leads to a highly negative $f^{\prime}$,
preventing the formation of attractor, the early-time attractor demands
a large enough $\Delta_{2}$, whereas the late-time attractor requires a smaller $\Delta_{2}$.

We also study the dependence of parameters $\Delta_{1}$ and $\alpha$ in Fig.~
\ref{fig:StreamlinesEarly}(c) and (d). Interestingly, we do not observe  attractor behavior when $\Delta_{1}$ and $\alpha$ are
large. It implies that the existence of the early-time attractor
in spin hydrodynamics demands small values for $\Delta_{1}$ and $\alpha$,
which is the same as the requirement of late-time attractor. This
observation is consistent with the analysis of Eq.~(\ref{eq:fprime_fw})
at the end of Sec. \ref{subsec:LateTime}.

Recalling definitions of $\alpha$ and $\Delta_{1,2}$ in Eq.~(\ref{eq:Kn-t}),
the above discussions indicate that if the dimensionless quantity
$\tau_{\phi}\gamma/\chi$ is smaller, the early-time attractor behavior will become more evident. 

\subsection{Discussion on small $\tau_{\phi}\gamma/\chi$ and $\Delta_{1}$}

As discussed in previous subsections, both the late-time and early-time attractors require that $\tau_{\phi}\gamma/\chi$ and $\Delta_{1}$ take small values. Now, we provide physical interpretation for this constraint. 

For simplicity, let us assume the spin susceptibility $\chi$ is $\mathcal{O}(\partial^0)$ and is approximately a constant in the following discussions. From Eqs.~(\ref{eq:Kn-t}), it is easy to get $\tau=\tau_1 w^{1/\Delta_1}$ and $\gamma=\alpha \chi \tau_1^{-1} w^{1+\Delta_2 - 1/\Delta_1}$. Within the dominan satisfying inequalities (\ref{eq:DomainSpinHydro}), we find that $\gamma$ decays as $w$ increases since $1+\Delta_2 - 1/\Delta_1$ is negative. Assuming that $\gamma$ has a small magnitude, the evolution equation of the source term $\phi^{xy}$ in Eq. (\ref{eq:Const2}) becomes
\begin{equation}
    \Delta_1 \frac{d\phi^{xy}}{dw}+\phi^{xy} = \mathcal{O}(\gamma),\label{eq:eqforphixy01}
\end{equation}
which gives 
\begin{equation}
    \phi^{xy}\approx\phi_{0}\exp\left(-\frac{w}{\Delta_{1}}\right)+\mathcal{O}(\gamma),\label{eq:solphixy01}
\end{equation}
with $\phi_{0}$ being a constant determined by initial conditions.
For a fixed $\Delta_{1}$, the spin source $\phi^{xy}$ decays exponentially
to $\mathcal{O}(\gamma)$, a small quantity as we assumed. 
When $\phi^{xy}\rightarrow0$ (i.e. $\gamma\rightarrow0$) at late times, the evolution equation for spin density
(\ref{eq:OriginalSpinEq}) is almost a conserved equation $\partial_{\lambda}\Sigma^{\lambda xy}\approx0$,
i.e., 
\begin{equation}
\frac{dS^{xy}}{d\tau}+\frac{1}{\tau}S^{xy}\approx0,
\end{equation}
which leads to\footnote{In Eq. (\ref{eq:Kn-t}), $\tau_{\phi}\gamma/\chi\ll1$ implies $\alpha\ll1$.
Thus, in the special case where $\Delta_{2}=-1$, we have $f\rightarrow-1-8\alpha\approx-1$
as $w\rightarrow\infty$. }, 
\begin{equation}
S^{xy}\approx S_{0}\frac{\tau_{1}}{\tau}=S_{0}w^{-1/\Delta_{1}},
\end{equation}
and then $f\approx-1$. Therefore, 
if $\gamma\rightarrow0$, the source term for the variation of spin, $\phi^{xy}$, 
decays exponentially to zero such that the spin can be approximated
as a conserved quantity, corresponding to $f\approx-1$, which explains
why we observe attractor in these cases. 

Next, we discuss the roles of $\tau_{\phi}$ and $\Delta_{1}$. From Eq.~(\ref{eq:eqforphixy01}), we observe that $\Delta_1$ acts as the effective 'relaxation time' for $\phi^{xy}(w)$ instead of $\tau_\phi$. When $\Delta_1$ decreases, $\phi^{xy}$ reaches its 'thermal equilibrium state', i.e., $\phi^{xy} \rightarrow 0$, more quickly. Since $\phi^{xy}$ is the source term that changes the spin, a smaller $\Delta_1$ makes $f$ converge to the attractors more easily.

\subsection{Singularities of $f$ }

\begin{figure}
\begin{centering}
\includegraphics[scale=0.85]{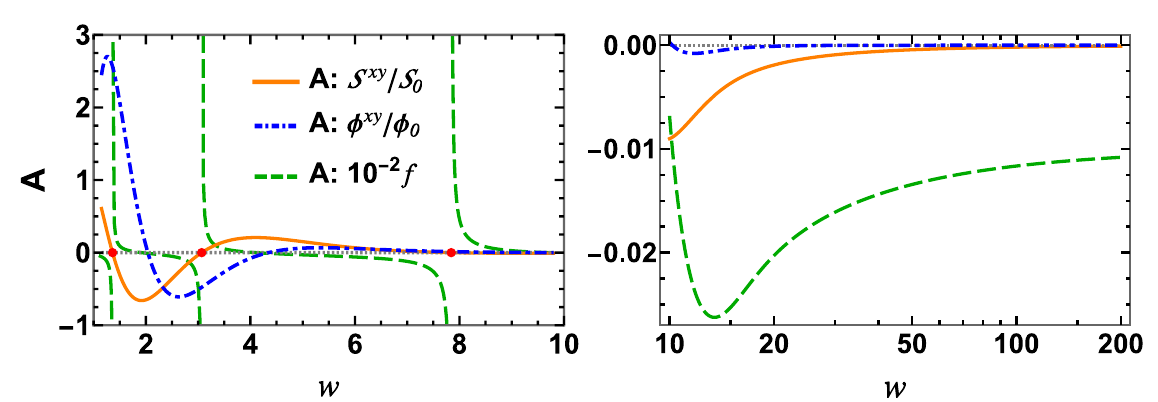}
\par\end{centering}
\caption{We plot a solution $S^{xy}(w)/S^{xy}(w_{0})$ and its corresponding $\phi^{xy}(w)/\phi^{xy}(w_0)$
and $10^{-2}f(w)$ as functions of $w$, where $w_{0}$ is the initial time.
The initial conditions are chosen as $w_{0}=1$ and $(S^{xy})^{\prime}(w_{0})=-2S^{xy}(w_{0})$.
The parameters are taken as $\Delta_{1}=1$, $\Delta_{2}=-2$, and
$\alpha=2$. The orange, blue dashdotted, and green dashed lines represent
$S^{xy}(w)/S^{xy}(w_{0})$, $\phi^{xy}(w)/\phi^{xy}(w_0)$, and $10^{-2}f(w)$, respectively.
The red points are the zeros of $S^{xy}(w)$. \label{fig:fSingular}}
\end{figure}

In Fig.~\ref{fig:StreamlinesLate} and \ref{fig:StreamlinesEarly},
we notice that some solutions of Eq.~(\ref{eq:RewriteSpinEvo1}) are
singular at finite $w$. Physically, these singular points of $f(w)$
correspond to the reversal of the sign of $S^{xy}$. Assuming that $S^{xy}=0$
and $dS^{xy}/dw\neq0$ at $w=w_{r}>0$, 
i.e., the spin sign is reversed at $w=w_{r}$,
Eq.~(\ref{eq:fS}) then gives,
\begin{eqnarray}
f(w_{r}-\epsilon) & = & \Delta_{1}\frac{w}{S^{xy}}\frac{dS^{xy}}{dw}\big|_{w=w_{r}-\epsilon}\approx\frac{\Delta_{1}w_{r}}{-\epsilon\frac{dS^{xy}}{dw}\big|_{w=w_{r}}}\frac{dS^{xy}}{dw}\big|_{w=w_{r}}=-\frac{\Delta_{1}w_{r}}{\epsilon},\label{eq:fmepsilon}\\
f(w_{r}+\epsilon) & = & \Delta_{1}\frac{w}{S^{xy}}\frac{dS^{xy}}{dw}\big|_{w=w_{r}+\epsilon}\approx\frac{\Delta_{1}w_{r}}{\epsilon\frac{dS^{xy}}{dw}\big|_{w=w_{r}}}\frac{dS^{xy}}{dw}\big|_{w=w_{r}}=\frac{\Delta_{1}w_{r}}{\epsilon},\label{eq:fpepsilon}
\end{eqnarray}
where $\epsilon$ is a constant obeying $0<\epsilon\ll w_{r}$. As
$\epsilon\rightarrow0^{+}$, we have $f(w_{r}-\epsilon)\rightarrow-\infty$
while $f(w_{r}+\epsilon)\rightarrow+\infty$ for $\Delta_{1}>0$,
indicating that $f(w)$ is singular at $w_{r}$. To provide a better
understanding, we plot $S^{xy}(w)$ and the corresponding $f(w)$
in Fig.~\ref{fig:fSingular}.
From the left panel of Fig.~\ref{fig:fSingular}, we find that the line for $S^{xy}(w)$
passes through the reds points, which represent the zeros of $S^{xy}(w)$, smoothly, but $f(w)$ is obviously singular at these points. 

Based on the above analysis, if $f(w)$ has several singularities,
then $S^{xy}(w)$ has different zeros, implying the oscillating behavior
of $S^{xy}(w)$ around $S^{xy}=0$. Comparing the left and right panels of 
Fig.~\ref{fig:fSingular}, we observe that, given initial conditions,
$f(w)$ can have many singularities at early-times, but $f(w)$ is regular
at late times due to the appearance of late-time
attractor. Thus the oscillating behavior of $S^{xy}(w)$ can be observed
at early times for appropriate parameter choices. 
The $\phi^{xy}(w)$, as the source term of the evolution equation
for spin density in Eq.~(\ref{eq:OriginalSpinEq}), may also
oscillate around $\phi^{xy}=0$ at early-times like $S^{xy}(w)$.
Such an oscillating behavior of $\phi^{xy}(w)$ is observed 
in the left panel of Fig.~\ref{fig:fSingular} at early-times.

\section{Focusing behavior of spin density \label{sec:FocusingBehavior}}

In this section, we report a new focusing behavior of spin density,
which has not been observed in the conventional relativistic hydrodynamics in a Bjorken flow.

The evolution equation for spin density, in Eq.~(\ref{eq:SpinDensityEq}) or Eq.~(\ref{eq:S_w}), is a second order linear ordinary differential
equation, whose general solution has the form of Eq.~(\ref{eq:LineaeComb}).
Imposing the initial conditions $S^{xy}(w_{0})=S_{0}$ and $(S^{xy})^{\prime}(w_{0})=S_{0}^{\prime}$
on Eq.~(\ref{eq:LineaeComb}) gives 
\begin{eqnarray}
c_{(1)} & = & \frac{1}{W_{0}}[S_{0}S_{(2)}^{\prime}(w_{0})-S_{0}^{\prime}S_{(2)}(w_{0})],\label{eq:CoefC1}\\
c_{(2)} & = & -\frac{1}{W_{0}}[S_{0}S_{(1)}^{\prime}(w_{0})-S_{0}^{\prime}S_{(1)}(w_{0})],\label{eq:CoefC2}
\end{eqnarray}
where $W_{0}$ is the Wronskian determinant,
\begin{equation}
W_{0}=\det\left(\begin{array}{cc}
S_{(1)}(w_{0}) & S_{(2)}(w_{0})\\
S_{(1)}^{\prime}(w_{0}) & S_{(2)}^{\prime}(w_{0})
\end{array}\right).
\end{equation}
The linear independence between $S_{(1)}$ and $S_{(2)}$ ensures
$W_{0}\neq0$ \citep{pontryagin:1962}. Substituting Eqs.~(\ref{eq:CoefC1},
\ref{eq:CoefC2}) into Eq.~(\ref{eq:LineaeComb}), the general solution
for $S^{xy}(w)$ can be written as  
\begin{eqnarray}
S^{xy}(w) & = & \frac{S_{0}}{W_{0}}\left[S_{(2)}^{\prime}(w_{0})S_{(1)}(w)-S_{(1)}^{\prime}(w_{0})S_{(2)}(w)\right]\nonumber \\
 &  & -\frac{S_{0}^{\prime}}{W_{0}}\left[S_{(2)}(w_{0})S_{(1)}(w)-S_{(1)}(w_{0})S_{(2)}(w)\right].\label{eq:SxyS0Sp0}
\end{eqnarray}
Here we note that $S^{xy}(w)$ depends linearly on $S_{0}$ and $S_{0}^{\prime}$. 

\begin{figure}
\begin{centering}
\includegraphics[scale=0.5]{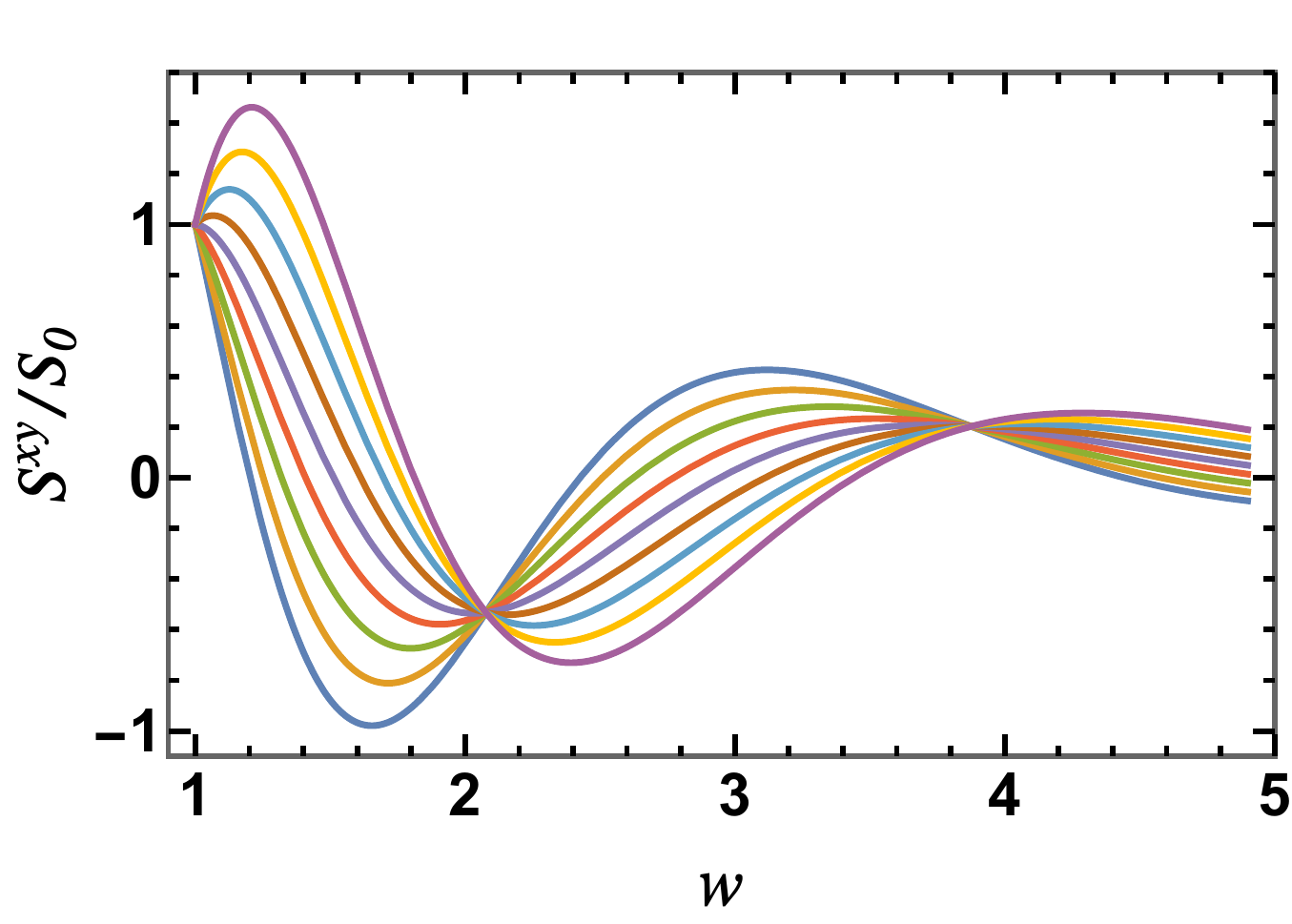}
\par\end{centering}
\caption{The focusing behavior for $S^{xy}(w)/S_{0}$ with different $S_{0}^{\prime}$.
The parameters are set to be $\Delta_{1}=1$, $\Delta_{2}=-1.5$,
and $\alpha=2$. The initial conditions are chosen as $w_{0}=1$ and
$S_{0}^{\prime}=$ $-4.9$, $-3.7$, $-2.5$, $-1.3$, $-0.1$, $1.1$,
$2.3$, $3.5$, and $4.7$. All solutions $S^{xy}(w)/S_{0}$ pass
through the same point at $w=2.077$, $3.876$, $6.804$, and $11.974$
(the last two are not shown in this figure). \label{fig:Sfocusing}}
\end{figure}

The structure of Eq.~(\ref{eq:SxyS0Sp0}) implies the focusing behavior of $S^{xy}(w)$. All solutions with the same $S_{0}$ but
different $S_{0}^{\prime}$ pass through some common points $(w_{c},S^{xy}(w_{c}))$, where $w_{c}\neq w_{0}$ is determined by solving 
\begin{equation}
    S_{(2)}(w_{0})S_{(1)}(w_c)=S_{(1)}(w_{0})S_{(2)}(w_c).\label{eq:FocusC01}
\end{equation}
That is because $S^{xy}(w_{c})$ is independent of $S_{0}^{\prime}$. 
Similarly,  the solutions
with the same $S_{0}^{\prime}$ but different $S_{0}$ focus on common
points $(w_{d},S^{xy}(w_{d}))$, where $w_{d}\neq w_{0}$ is determined by solving
\begin{equation}
    S_{(2)}^{\prime}(w_{0})S_{(1)}(w_d)=S_{(1)}^{\prime}(w_{0})S_{(2)}(w_d),\label{eq:FocusC02}
\end{equation}
We plot several solutions $S^{xy}(w)/S_{0}$
with different $S_{0}^{\prime}$ in Fig.~\ref{fig:Sfocusing} as
an example. 
It is clear to see the focusing behavior of spin density, like a beam of light passing through a convex lens to focus on a point.


Let us comment on the focusing behavior. First, the focusing
behavior of $S^{xy}(w)$ arises from the fact that the spin
evolution equation (\ref{eq:S_w}) is a linear differential equation.
If we add nonlinear terms to Eq.~(\ref{eq:S_w}), the dependence of
the solution on the initial conditions may be nonlinear unlike Eq.~(\ref{eq:SxyS0Sp0}),
and the focusing behavior may disappear. Second, the focusing behavior
disappears for some inappropriate choices of $w_{0}$, $\alpha$,
and $\Delta_{1,2}$, since Eqs.~(\ref{eq:FocusC01}, \ref{eq:FocusC02})
have no nontrivial solutions in such cases. Third, the common points
depend on the values of $w_{0}$, $\alpha$,
and $\Delta_{1,2}$. Finally, numerical results reveal that there
are only finitely many common points, i.e., we do not observe the focusing behavior for sufficiently large $w$.

\section{Conclusion \label{sec:Conclusion-} }

In this work, we have investigated the late-time asymptotic solutions, attractor, and focusing behavior of the minimal causal spin hydrodynamics in a Bjorken flow, which is  boost invariant in longitudinal direction and homogeneous in transverse plane.
Following our previous work \citep{Wang:2021ngp},
we derive the evolution equations for energy density 
and spin density in Eqs.~(\ref{eq:EnergyDensityEq})
and (\ref{eq:SpinDensityEq}), respectively.

To study the time evolution of spin density, we introduce the Knudsen number in Eq.~(\ref{eq:Knudsen_number}) and parameterize
it in Eq.~(\ref{eq:Kn-t}), corresponding to the cases where Knudsen number is time-dependent. 
By using the method of dominant balance, we obtain
the asymptotic solutions of spin density $S^{xy}$ at late times, shown in Table.~\ref{tab:AsympSolLarge}.
We find that the spin density can be power-law decaying, 
if the parameters $\Delta_{1,2}$ satisfy inequalities (\ref{eq:DomainSpinHydro}).

Within the domain in which the inequalities (\ref{eq:DomainSpinHydro}) are satisfied,
we observe both late-time and early-time attractors for 
the decay rate of spin density $f$. 
The attractor behavior becomes more
evident, when the dimensionless quantities 
$\tau_{\phi}\gamma/\chi$ and $\Delta_{1}$ are smaller. 
We also discuss the singularities of $f$, which arise from the definition of $f$
and correspond to the reversal of spin sign. At early-times, $f$ can have many singularities,
implying the oscillating behavior of spin density and spin source
term.

We also report the focusing behavior in spin hydrodynamics which has not been found in the conventional relativistic hydrodynamics in Bjorken expansion.
Specifically, the solutions with the same $S_{0}$ but
different $S_{0}^{\prime}$, or the same $S_{0}^{\prime}$ but different
$S_{0}$, can pass through some common points, much like a beam of
light passing through a convex lens to focus on a point.

Remarkably, the finding of power-law decaying and attractors in spin hydrodynamics indicates that the spin density can be treated as a conventional hydrodynamic variable at late times under certain conditions. 

\begin{acknowledgments}
This work is supported in part by the National Key Research and Development
Program of China under Contract No. 2022YFA1605500, by the Chinese
Academy of Sciences (CAS) under Grant No. YSBR-088 and by National Natural Science Foundation of China (NSFC) under Grants No. 11975079, 12075235 and No. 12135011.
\end{acknowledgments}

\appendix

\section{Equations in Bjorken expansion \label{sec:Equations-in-Bjorken} }

We present a detailed derivation of the energy
density equation (\ref{eq:EnergyDensityEq}) and spin density equation
(\ref{eq:SpinDensityEq}). Considering the Bjorken velocity profile in Eq.~(\ref{eq:BjorkenVelo}),
the conservation equations $\nabla_{\mu}\Theta^{\mu\nu}=0$ and $\nabla_{\lambda}J^{\lambda\mu\nu}=0$
demand that   
\begin{eqnarray}
0 & = & \Delta_{\rho\nu}\nabla_{\mu}\pi^{\mu\nu}-\Delta_{\nu\rho}\partial_{\tau}q^{\nu}-\frac{1}{\tau}q_{\rho}+\Delta_{\rho\nu}\nabla_{\mu}\phi^{\mu\nu},\label{eq:Conser1}\\
0 & = & \partial_{\tau}e+(e+p+\Pi)\frac{1}{\tau}-\nabla_{\mu}u_{\nu}\pi^{\mu\nu}+\nabla_{\mu}q^{\mu}-\nabla_{\mu}u_{\nu}\phi^{\mu\nu},\\
0 & = & \partial_{\tau}S^{\mu\tau}+\frac{1}{\tau}S^{\mu\tau}+\Gamma_{\tau\xi}^{\mu}S^{\xi\tau}+2q^{\mu},\\
0 & = & \Delta_{\mu}^{\rho}\Delta_{\nu}^{\sigma}\partial_{\tau}S^{\mu\nu}+\Delta_{\mu}^{\rho}\Delta_{\nu}^{\sigma}\Gamma_{\xi\tau}^{\xi}S^{\mu\nu}+\Delta_{\mu}^{\rho}\Delta_{\nu}^{\sigma}\Gamma_{\tau\xi}^{\mu}S^{\xi\nu}+\Delta_{\mu}^{\rho}\Delta_{\nu}^{\sigma}\Gamma_{\tau\xi}^{\nu}S^{\mu\xi}+2\phi^{\rho\sigma},\label{eq:Conser4}
\end{eqnarray}
where $\Gamma_{\mu\nu}^{\lambda}$ represent the Christoffel symbols.
Let us now suppose that, in the $(\tau,x,y,\xi)$ coordinates, the
quantities $e$, $p$, $\Pi$, $\pi^{\mu\nu}$, $q^{\mu}$, and $S^{\mu\nu}$
are functions of $\tau$ only. Then, the Eqs.~(\ref{eq:Conser1} - \ref{eq:Conser4})
can be simplified as follows,   
\begin{eqnarray}
\frac{d}{d\tau}(\tau q^{x}) & = & 0,\label{eq:EomVel1}\\
\frac{d}{d\tau}(\tau q^{y}) & = & 0,\\
\frac{d}{d\tau}(\tau q^{\xi}) & = & 0,\label{eq:EomVel3}\\
\frac{d}{d\tau}S^{x\tau}+\frac{1}{\tau}S^{x\tau}+2q^{x} & = & 0,\label{eq:EomSpin1}\\
\frac{d}{d\tau}S^{y\tau}+\frac{1}{\tau}S^{y\tau}+2q^{y} & = & 0,\\
\frac{d}{d\tau}(\tau S^{\xi\tau})+\frac{1}{\tau}(\tau S^{\xi\tau})+2\tau q^{\xi} & = & 0,\label{eq:EomSpin3}\\
\nonumber \\
\frac{d}{d\tau}e+(e+p+\Pi)\frac{1}{\tau}+\tau\pi^{\xi\xi} & = & 0,\label{eq:EqEnergyDensity}\\
\nonumber \\
\frac{d}{d\tau}S^{xy}+\frac{1}{\tau}S^{xy}+2\phi^{xy} & = & 0,\label{eq:eqSxy}\\
\frac{d}{d\tau}S^{x\xi}+\frac{2}{\tau}S^{x\xi}+2\phi^{x\xi} & = & 0,\label{eq:eqSxxi}\\
\frac{d}{d\tau}S^{y\xi}+\frac{2}{\tau}S^{y\xi}+2\phi^{y\xi} & = & 0.\label{eq:eqSyxi}
\end{eqnarray}
Similarly, in Bjorken expansion, the constitutive equations (\ref{eq:Const1} - \ref{eq:Const4})
can be written as follows,    
\begin{eqnarray}
\tau_{\Pi}\frac{d}{d\tau}\Pi+\Pi & = & -\frac{\zeta}{\tau},\label{eq:ConstB1}\\
\nonumber \\
\tau_{\pi}\frac{d}{d\tau}\pi^{xy}+\pi^{xy} & = & 0,\\
\tau_{\pi}\frac{d}{d\tau}(\tau\pi^{x\xi})+\tau\pi^{x\xi} & = & 0,\\
\tau_{\pi}\frac{d}{d\tau}(\tau\pi^{y\xi})+\tau\pi^{y\xi} & = & 0,\\
\tau_{\pi}\frac{d}{d\tau}\pi^{xx}+\pi^{xx} & = & \frac{2\eta}{3\tau},\\
\tau_{\pi}\frac{d}{d\tau}(\tau^{2}\pi^{\xi\xi})+\tau^{2}\pi^{\xi\xi} & = & -\frac{4\eta}{3\tau},\\
\nonumber \\
\tau_{\phi}\frac{d}{d\tau}\phi^{xy}+\phi^{xy} & = & \frac{4\gamma}{\chi}S^{xy},\label{eq:eqphixy}\\
\tau_{\phi}\frac{d}{d\tau}(\tau\phi^{x\xi})+\tau\phi^{x\xi} & = & \frac{4\gamma}{\chi}\tau S^{x\xi},\label{eq:eqphixxi}\\
\tau_{\phi}\frac{d}{d\tau}(\tau\phi^{y\xi})+\tau\phi^{y\xi} & = & \frac{4\gamma}{\chi}\tau S^{y\xi},\label{eq:eqphiyxi}\\
\nonumber \\
\tau_{q}\frac{d}{d\tau}q^{x}+q^{x} & = & \frac{4\lambda}{\chi-\chi^{\prime}}S^{\tau x},\label{eq:ConstB10}\\
\tau_{q}\frac{d}{d\tau}q^{y}+q^{y} & = & \frac{4\lambda}{\chi-\chi^{\prime}}S^{\tau y},\\
\tau_{q}\frac{d}{d\tau}(\tau q^{\xi})+\tau q^{\xi} & = & \frac{4\lambda}{\chi-\chi^{\prime}}\tau S^{\tau\xi}.\label{eq:ConstB12}
\end{eqnarray}
Here we have used the equations of state (\ref{eq:Eos2}) to replaced
$\omega^{\mu\nu}$ by $S^{\mu\nu}$.

It is obvious that there are $9$ equations, i.e., Eqs.~(\ref{eq:EomVel1}-\ref{eq:EomSpin3})
and Eqs.~(\ref{eq:ConstB10}-\ref{eq:ConstB12}), corresponding to the
$6$ variables $q^{x},q^{y},q^{\xi}$, and $S^{x\tau},S^{y\tau},S^{\xi\tau}$.
One way to get a self-consistent solution is to set
\begin{equation}
q^{x},q^{y},q^{\xi},S^{x\tau},S^{y\tau},S^{\xi\tau}=0,
\end{equation}
at initial proper time. Then they are always zero at any later time according to
the Eqs.~(\ref{eq:EomVel1}-\ref{eq:EomSpin3}) and (\ref{eq:ConstB10}-\ref{eq:ConstB12}). 

Subsequently, we are left with $13$ equations associated with $13$
unknowns, namely $e$, $\Pi$, $\pi^{xx}$, $\pi^{xy}$, $\pi^{x\xi}$,
$\pi^{y\xi}$, $\pi^{\xi\xi}$, $\phi^{xy}$, $\phi^{x\xi}$, $\phi^{y\xi}$,
$S^{xy}$, $S^{x\xi}$, and  $S^{y\xi}$. The energy density equation
is given by Eq.~(\ref{eq:EqEnergyDensity}), the same as those in conventional
hydrodynamics. Taking $\pi^{\mu\nu},\Pi=0$ and Eq.~(\ref{eq:Eos2}), Eq.~(\ref{eq:EqEnergyDensity})
reduces to Eq.~(\ref{eq:EnergyDensityEq}). The evolution equation of spin density
(\ref{eq:SpinDensityEq}) is derived by eliminating the variable $\phi^{\mu\nu}$
in Eqs.~(\ref{eq:eqSxy}-\ref{eq:eqSyxi}) and Eqs.~(\ref{eq:eqphixy}-\ref{eq:eqphiyxi}).
We emphasize that the simplification $\pi^{\mu\nu},\Pi=0$ does not
alter the evolution equation of spin density, because Eqs.~(\ref{eq:eqSxy}-\ref{eq:eqSyxi})
and Eqs.~(\ref{eq:eqphixy}-\ref{eq:eqphiyxi}) are independent of
$\pi^{\mu\nu}$ and $\Pi$.

\section{Evolution equation for spin density with a constant Knudsen number \label{sec:AsymptoticS_C}}

We study specific scenarios where the
Knudsen number remains constant over time. Since $w=\tau/\tau_{\phi}$
becomes a constant, the parameterization in Eq.~(\ref{eq:Kn-t}) is inappropriate,
preventing us from expressing Eq.~(\ref{eq:SpinDensityEq}) in the
form of Eq.~(\ref{eq:S_w}). To address this issue, we introduce the
initial time $\tau_{0}>0$ and assume that
\begin{equation}
\tau_{\phi}\equiv c_{\phi}\tau,\ v\equiv\frac{\tau}{\tau_{0}},\ \frac{\tau_{\phi}\gamma}{\chi}=\beta v^{\Delta_{3}},\label{eq:Kn-c}
\end{equation}
where $c_{\phi}>0$ serves as the Knudsen number, $\beta>0$ and $\Delta_{3}$
are dimensionless constants. According to the dimensional analysis in Eq.~(\ref{eq:DimensionAna01}), the parameterization in Eq.~(\ref{eq:Kn-c})
demands that the system is massive. 

\subsection{Late-time asymptotic solutions \label{subsec:AsymptoticSol_gv}}

Taking the parameterization in Eq.~(\ref{eq:Kn-c}), the evolution equation for spin
(\ref{eq:SpinDensityEq}) can be written as 
\begin{equation}
\frac{d^{2}S^{xy}}{dv^{2}}+(c_{\phi}^{-1}+1)v^{-1}\frac{dS^{xy}}{dv}+v^{-2}(c_{\phi}^{-1}-1+8\beta c_{\phi}^{-2}v^{\Delta_{3}})S^{xy}=0.\label{eq:ConstKn_SpinEvo}
\end{equation}
Similar to Eq.~(\ref{eq:fS}), we introduce an auxiliary function,
\begin{equation}
g(v)\equiv\frac{v}{S^{xy}}\frac{dS^{xy}}{dv}=\frac{\tau}{S^{xy}}\frac{dS^{xy}}{d\tau},\label{eq:gv}
\end{equation}
such that Eq.~(\ref{eq:ConstKn_SpinEvo}) is reduced to a first order ordinary
differential equation 
\begin{equation}
vg^{\prime}+g^{2}+c_{\phi}^{-1}g+c_{\phi}^{-1}-1+8\beta c_{\phi}^{-2}v^{\Delta_{3}}=0,\label{eq:Eq_gv}
\end{equation}
where $g^{\prime}=dg/dv$. 

To analyze the long-time evolution of spin density, we compute the
asymptotic solutions of Eq.~(\ref{eq:ConstKn_SpinEvo}). Employing the method of dominant balance \citep{bender1999advanced,paulsen2013asymptotic},
we first derive the asymptotic solutions of Eq.~(\ref{eq:Eq_gv}) as
$v\rightarrow+\infty$ (i.e., $\tau\rightarrow+\infty$). As shown
in Sec.  \ref{sec:SolveSpinEq}, the solutions of Eq.~(\ref{eq:ConstKn_SpinEvo})
can be constructed from any two special  solutions of Eq.~(\ref{eq:Eq_gv})
by integrating Eq.~(\ref{eq:gv}). The final results are categorized
based on different values of $\Delta_{3}$, and shown in Table.~\ref{tab:ConstKn_LargeSol}. 

\begin{table}
\caption{Asymptotic solutions of Eq.~(\ref{eq:ConstKn_SpinEvo}) as $v\rightarrow+\infty$
(i.e., $\tau\rightarrow+\infty$). Here $S_{(1)}$ and $S_{(2)}$
represent two linearly independent solutions of Eq.~(\ref{eq:ConstKn_SpinEvo}).
In the case of $\Delta_{3}=0$, we have used the shorthand notation
$\Gamma\equiv(2c_{\phi}-1)^{2}-32\beta$. \label{tab:ConstKn_LargeSol}}

\noindent \centering{}%
\begin{tabular}{c|ccc}
\cline{1-3} \cline{2-3} \cline{3-3} 
\noalign{\vskip0.2cm}
\multicolumn{3}{c}{$v\rightarrow+\infty$} & \tabularnewline[0.2cm]
\cline{1-3} \cline{2-3} \cline{3-3} 
\multirow{4}{*}{$\Delta_{3}>0$} &  & \multirow{2}{*}{$S_{(1)}\sim v^{-(\Delta_{3}+2c_{\phi}^{-1})/4}\cos\left(\frac{4\sqrt{2\beta}v^{\Delta_{3}/2}}{c_{\phi}\Delta_{3}}\right)$,} & \multirow{2}{*}{}\tabularnewline
 &  &  & \tabularnewline
 &  & \multirow{2}{*}{$S_{(2)}\sim v^{-(\Delta_{3}+2c_{\phi}^{-1})/4}\sin\left(\frac{4\sqrt{2\beta}v^{\Delta_{3}/2}}{c_{\phi}\Delta_{3}}\right)$.} & \tabularnewline
 &  &  & \tabularnewline
\cline{1-3} \cline{2-3} \cline{3-3} 
\multirow{6}{*}{$\Delta_{3}=0$} &  & \multirow{2}{*}{$S_{(1)}=v^{-\frac{1}{2c_{\phi}}(1+\sqrt{\Gamma})}$, $S_{(2)}=v^{-\frac{1}{2c_{\phi}}(1-\sqrt{\Gamma})}$
for $\Gamma>0$,} & \tabularnewline
 &  &  & \tabularnewline
 &  & \multirow{2}{*}{$S_{(1)}=v^{-\frac{1}{2c_{\phi}}}$, $S_{(2)}=v^{-\frac{1}{2c_{\phi}}}\ln v$
for $\Gamma=0$,} & \tabularnewline
 &  &  & \tabularnewline
 &  & \multirow{2}{*}{$S_{(1)}=v^{-\frac{1}{2c_{\phi}}}\cos\left(\frac{\sqrt{-\Gamma}\ln v}{2c_{\phi}}\right)$,
$S_{(2)}=v^{-\frac{1}{2c_{\phi}}}\sin\left(\frac{\sqrt{-\Gamma}\ln v}{2c_{\phi}}\right)$
for $\Gamma<0$.} & \tabularnewline
 &  &  & \tabularnewline
\cline{1-3} \cline{2-3} \cline{3-3} 
\multirow{4}{*}{$\Delta_{3}<0$} &  & \multirow{2}{*}{$S_{(1)}\sim v^{-1}$, $S_{(2)}\sim v^{1-c_{\phi}^{-1}}$ for $c_{\phi}\neq\frac{1}{2}$,} & \tabularnewline
 &  &  & \tabularnewline
 &  & \multirow{2}{*}{$S_{(1)}\sim v^{-1}$, $S_{(2)}\sim v^{-1}\ln v$ for $c_{\phi}=\frac{1}{2}$.} & \tabularnewline
 &  &  & \tabularnewline
\cline{1-3} \cline{2-3} \cline{3-3} 
\end{tabular}
\end{table}

Since $c_{\phi}$ plays the role of Knudsen number, i.e., $c_{\phi}=\tau_{\phi}/\tau\approx\mathrm{Kn}$,
the domain of spin hydrodynamics lies within $0<c_{\phi}\leq1$. From
Table.~\ref{tab:ConstKn_LargeSol}, we find that, for $0<c_{\phi}\leq1$,
the spin density exhibits power-law decay (with or without oscillation),
instead of exponential decay. Thus the spin density can be regarded
as a hydrodynamic variable, and the spin effects cannot be lightly
ignored. On the contrary, the spin density can grow to infinity if
$c_{\phi}>1$, e.g., we have $S_{(2)}\sim v^{1-c_{\phi}^{-1}}$ in
the case of $\Delta_{3}<0$, $c_{\phi}\neq\frac{1}{2}$. 
Since it is impossible for real systems to contain infinite spin 
density, spin hydrodynamics loses its applicability if
$c_{\phi}>1$, in line with the interpretation of $c_{\phi}$ as the
Knudsen number. 

\subsection{Slow convergence to late-time attractor }

\begin{figure}
\noindent \begin{centering}
\includegraphics[scale=0.6]{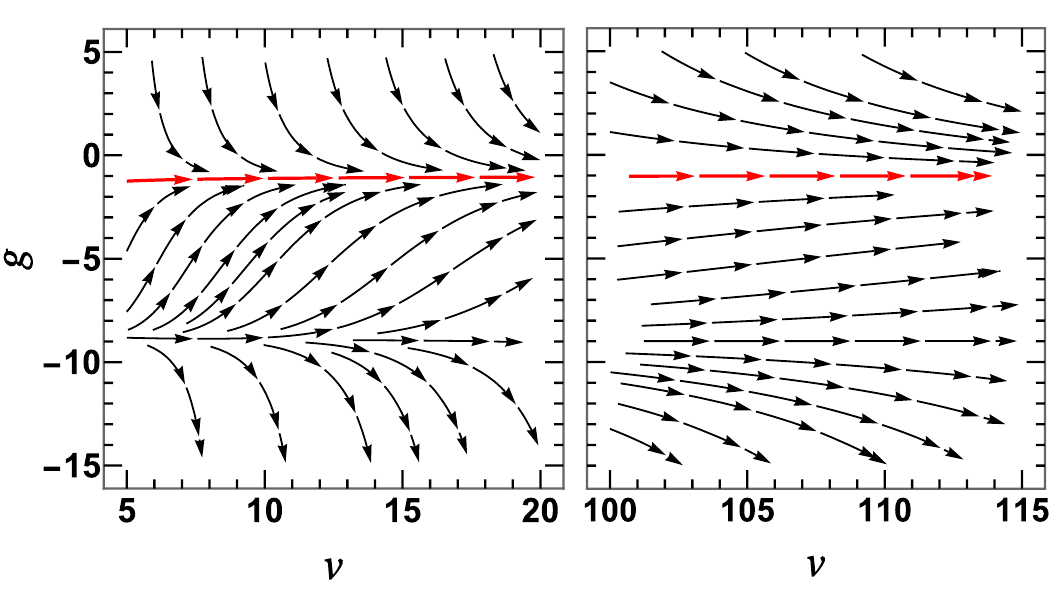}
\par\end{centering}
\caption{Streamlines for the equation (\ref{eq:Eq_gv}) are plotted when $5<v<20$
and $100<v<115$ with $\beta=0.01$, $c_{\phi}=0.1$, and $\Delta_{3}=-1$.
The red line stands for the attractor. \label{fig:gv_latetime}}
\end{figure}

Based on fixed point analysis in Refs.~\citep{Blaizot:2021cdv}, we find that
there exists a late-time attractor under certain conditions. As $v\rightarrow+\infty$,
we have stable fixed points 

\begin{equation}
g\rightarrow\max\{-1,1-c_{\phi}^{-1}\}\quad\textrm{for }\Delta_{3}<0,c_{\phi}\neq\frac{1}{2},\label{eq:gv_fix_1}
\end{equation}
and 
\begin{equation}
g\rightarrow\frac{-1+\sqrt{\Gamma}}{2c_{\phi}}\quad\textrm{for }\Delta_{3}=0,\Gamma>0,\label{eq:gv_fix_2}
\end{equation}
where $\Gamma\equiv(2c_{\phi}-1)^{2}-32\beta$. When considering other
values for $c_{\phi}$, $\beta$, and $\Delta_{3}$, it is observed
that no stable fixed point emerges as $v\rightarrow+\infty$. The
late-time attractor is represented in the streamlines
of Eq.~(\ref{eq:Eq_gv}), as shown in the left panel of Figure. \ref{fig:gv_latetime}. 

We write Eq.~(\ref{eq:Eq_gv}) as 
\begin{equation}
g^{\prime}=-\frac{1}{v}(g^{2}+c_{\phi}^{-1}g+c_{\phi}^{-1}-1+8\beta c_{\phi}^{-2}v^{\Delta_{3}}).\label{eq:Dgv}
\end{equation}
Analogous to the analysis in Sec. \ref{subsec:LateTime}, we find
that the parameters $c_{\phi}$, $\beta$, $\Delta_{3}$ in Eq.~(\ref{eq:Dgv})
play roles similar to those of $\Delta_{1}$, $\alpha$, $\Delta_{2}$
in Eq.~(\ref{eq:fprime_fw}), respectively. It means that the streamlines
around the late-time attractor, shown in Eqs.~(\ref{eq:gv_fix_1}, \ref{eq:gv_fix_2}),
will converge rapidly to the attractor for small $c_{\phi}$, $\beta$,
and $\Delta_{3}$. 

In Eq.~(\ref{eq:Dgv}), it is worth noting that 
\begin{equation}
g^{\prime}\rightarrow0\ \textrm{ as }v\rightarrow+\infty,
\end{equation}
for any $\Delta_{3}\leq0$ and finite $g$, which indicates that the
speed of the convergence to the late-time attractor becomes slower
as $v$ increases. Comparing the left panel and right panel in Figure.
\ref{fig:gv_latetime}, the convergence to the attractor at $v=110$
is much slower than that at $v=10$. Notably, this would not occur
in the cases discussed in Sec. \ref{subsec:LateTime}, since $f^{\prime}$
tends to a nonzero constant as $w\rightarrow+\infty$, i.e., the convergent
speed is nonzero at late times.

\section{Mathematical structure of the transseries solutions of Eq.~(\ref{eq:RewriteSpinEvo1})
\label{sec:TransseriesSol}}

We discuss the mathematical structure of the transseries solutions
of Eq.~(\ref{eq:RewriteSpinEvo1}), as $w\rightarrow+\infty$. According
to the physical requirements in Sec. \ref{sec:SolveSpinEq}, we focus
on the cases where $\Delta_{1}>0$ and $\Delta_{2}\leq-1$. 

\subsection{$\Delta_{2}$ is an integer}

The simplest case is that $\Delta_{2}$ is an integer, i.e., $\Delta_{2}\in\{-1,-2,...\}$.
Taking the transformation 
\begin{equation}
f=\begin{cases}
h-(1+8\alpha), & \textrm{for }\Delta_{2}=-1,\\
h-1, & \textrm{for }\Delta_{2}\in\{-2,-3,...\},
\end{cases}
\end{equation}
Eq.~(\ref{eq:RewriteSpinEvo1}) can be equivalently rewritten as the
form  
\begin{equation}
h^{\prime}=-\Delta_{1}^{-1}h+\sigma w^{-1}h+I(w,h),\label{eq:PreparedForm}
\end{equation}
where 
\begin{eqnarray}
\sigma & \equiv & \begin{cases}
2(1+8\alpha)\Delta_{1}^{-1}, & \textrm{for }\Delta_{2}=-1,\\
2\Delta_{1}^{-1}, & \textrm{for }\Delta_{2}\in\{-2,-3,...\},
\end{cases}\\
I(x,y) & \equiv & \begin{cases}
[(1+8\alpha)^{2}-1]\Delta_{1}^{-1}w^{-1}+\Delta_{1}^{-1}w^{-1}h^{2}, & \textrm{for }\Delta_{2}=-1,\\
8\alpha\Delta_{1}^{-1}w^{1+\Delta_{2}}+\Delta_{1}^{-1}w^{-1}h^{2}, & \textrm{for }\Delta_{2}\in\{-2,-3,...\}.
\end{cases}
\end{eqnarray}
This form in Eq.~(\ref{eq:PreparedForm}) is called prepared form\footnote{The rigorous definition of prepared form in Ref.~\citep{costin2008asymptotics}
has more constraints on $I(x,y)$, but those constraints can be easily
fulfilled by redefining $h$ in Eq.~(\ref{eq:PreparedForm}). More discussions
can be found in Chapter 5.6 of Ref.~\citep{costin2008asymptotics}.} \citep{costin2008asymptotics}. The general transseries solution
of Eq.~(\ref{eq:PreparedForm}) is given by \citep{costin2008asymptotics}
\begin{equation}
h=\sum_{k=0}^{+\infty}C^{k}e^{-k\Delta_{1}^{-1}w}w^{k\sigma}h_{k},\label{eq:Transseriesh}
\end{equation}
where $C$ is an undetermined constant, and $h_{k}$ are formal power
series taking the form, 
\begin{equation}
h_{k}=\sum_{l=1}^{+\infty}a_{k,l}w^{-l},\quad k=0,1,2,....\label{eq:hk}
\end{equation}
Substituting Eqs.~(\ref{eq:Transseriesh}, \ref{eq:hk}) into Eq.~(\ref{eq:PreparedForm}),
we can calculate the coefficients $a_{k,l}$ order by order \citep{costin2008asymptotics}. 

The initial information is encoded in the constant $C$ in Eq.~(\ref{eq:Transseriesh}).
Note that the constant $C$ always appears as $Ce^{-\Delta_{1}^{-1}w}$.
As $w\rightarrow+\infty$, the $Ce^{-\Delta_{1}^{-1}w}$
decays faster than any power series. Hence, solutions with different
$C$ converge to $h_{0}$, resulting in the late-time attractor behavior.
In other words, the information of initial state is lost at late times. 

\subsection{$\Delta_{2}$ is a non-integer rational number \label{subsec:rationalnumber}}

If $\Delta_{2}$ is a non-integer rational number less than $-1$,
there exists two positive integers $p$, $q$ such that 
\begin{equation}
1+\Delta_{2}=-\frac{p}{q},
\end{equation}
where $q\geq2$ and $p/q$ is irreducible. Let 
\begin{equation}
f=h-1.
\end{equation}
The equation for $h$ is given by 
\begin{equation}
h^{\prime}+\Delta_{1}^{-1}h-2\Delta_{1}^{-1}w^{-1}h+8\alpha\Delta_{1}^{-1}w^{-p/q}+\Delta_{1}^{-1}w^{-1}h^{2}=0.\label{eq:Delta2NonIntEqh}
\end{equation}
Since $p/q$ is a non-integer, one cannot transform Eq.~(\ref{eq:Delta2NonIntEqh})
into the prepared form (\ref{eq:PreparedForm}) 

Similar to Eq.~(\ref{eq:Transseriesh}), we find that the solution
of Eq.~(\ref{eq:Delta2NonIntEqh}) has the structure 
\begin{equation}
h=\sum_{k=0}^{+\infty}C^{k}e^{-k\Delta_{1}^{-1}w}w^{2k\Delta_{1}^{-1}}h_{k},\label{eq:TransserieSol2}
\end{equation}
but here $h_{k}$ is a formal power series of $w^{1/q}$ instead of
$w$, 
\begin{equation}
h_{k}=\sum_{l=1}^{+\infty}b_{k,l}w^{-l/q}.\label{eq:TransseriesSolhk2}
\end{equation}
Again, in Eq.~(\ref{eq:TransserieSol2}), we see that the constant
$C$ encoding initial information always times a the exponential decaying
factor $e^{-\Delta_{1}^{-1}w}$. 

\subsection{$\Delta_{2}$ is an irrational number}

If $\Delta_{2}$ is an irrational number, the mathematical structure
of the transseries solution of Eq.~(\ref{eq:RewriteSpinEvo1}) becomes
complicated. However, we can use a rational number to approximate
the irrational number $\Delta_{2}$, such that the approximated solution
resembles Eq.~(\ref{eq:TransserieSol2}). This is because solutions
of Eq.~(\ref{eq:RewriteSpinEvo1}) are continuous functions of $\Delta_{2}$. 

Let us fist give a brief explanation for the continuous dependence
of solutions of Eq.~(\ref{eq:S_w}) on parameter $\Delta_{2}$. The
second order linear equation (\ref{eq:S_w}) can be transformed into
the first order linear equations 
\begin{equation}
\frac{d}{dw}\left(\begin{array}{c}
S^{xy}\\
(S^{xy})^{\prime}
\end{array}\right)=-\mathrm{A}(w;\alpha,\Delta_{1},\Delta_{2})\left(\begin{array}{c}
S^{xy}\\
(S^{xy})^{\prime}
\end{array}\right),\label{eq:FirstOrderLinearS}
\end{equation}
where $(S^{xy})^{\prime}\equiv dS^{xy}/dw$, and 
\begin{equation}
\mathrm{A}(w;\alpha,\Delta_{1},\Delta_{2})\equiv\left(\begin{array}{cc}
0 & -1\\
\Delta_{1}^{-2}(w^{-1}-w^{-2}+8\alpha w^{\Delta_{2}}) & \Delta_{1}^{-1}+w^{-1}
\end{array}\right).
\end{equation}
For simplicity, let us fix $\alpha>0$ and $\Delta_{1}>0$. The function
$\mathrm{A}(w;\alpha,\Delta_{1},\Delta_{2})$ is well-defined and
continuous for all $w>0$, and $\Delta_{2}<-1$. According to Theorem
13 of Ref.~\citep{pontryagin:1962}, for arbitrary numbers $w_{(0)}>0$
and $\Delta_{2(0)}<-1$, there exist positive numbers $L_{0}$ and
$\epsilon$, such that solutions of Eq.~(\ref{eq:FirstOrderLinearS})
are continuous on the interval $|w-w_{(0)}|<L_{0}$ and $|\Delta_{2}-\Delta_{2(0)}|<\epsilon$.
Hence, $S^{xy}$ and $(S^{xy})^{\prime}$ are continuous functions
of $w$ and $\Delta_{2}$ on such interval. 

Recall that the solutions of Eq.~(\ref{eq:RewriteSpinEvo1}) can be
constructed from $f=w(S^{xy})^{\prime}/S^{xy}$, with $S^{xy}\neq0$.
The function $f$ must be continuous for $\Delta_{2}<-1$. When $\Delta_{2}$
is an irrational number, we can decompose it into 
\begin{equation}
\Delta_{2}=r+\delta,
\end{equation}
where $r$ is a rational number, while $\delta$ is irrational. Then
we replace $\Delta_{2}$ by $r$ in Eq.~(\ref{eq:RewriteSpinEvo1}).
Because of the continuity, the difference between the approximated
solutions (with $\Delta_{2}\rightarrow r$) and the exact solution
tends to zero when $\delta\rightarrow0$.

\bibliographystyle{h-physrev}
\bibliography{refs}

\end{document}